\documentclass[pra, notitlepage, superscriptaddress,  reprint]{revtex4-1}
\usepackage{amssymb}
\usepackage{amsmath}
\usepackage{braket}
\usepackage{graphicx}
\usepackage[usenames,dvipsnames]{color}
\usepackage{tensor}
\usepackage{empheq}
\usepackage{capt-of}
\usepackage[normalem]{ulem} 
\usepackage{soul} %

\definecolor{myorange}{RGB}{199.24, 87.48, 47.80}
\usepackage[colorlinks,bookmarks=false,citecolor=NavyBlue,linkcolor=OliveGreen,urlcolor=blue]{hyperref}
\newcommand{\be}{\begin{equation}}
\newcommand{\ee}{\end{equation}}
\newcommand{\ba}{\begin{aligned}}
\newcommand{\ea}{\end{aligned}}
\newcommand{\bw}{\begin{widetext}}
\newcommand{\ew}{\end{widetext}}
\newcommand{\xt}{\zeta}
\newcommand{\T}{\hat{ T}}
\newcommand{\sg}{\hat{ \sigma}}

\newcommand{\1}{\hat{ 1}}

\newcommand{\tr}[2]{\mathrm{tr}_{#1}(#2)}

\def\epp{\: .}
\def\epc{\: ,}

\def\doi{http://dx.doi.org/}

\begin{document}
\title{Transport in Out-of-Equilibrium XXZ Chains: Exact Profiles of Charges and Currents}
\author{Bruno Bertini}
\affiliation{SISSA and INFN, via Bonomea 265, 34136, Trieste, Italy}
\author{Mario Collura}
\affiliation{SISSA and INFN, via Bonomea 265, 34136, Trieste, Italy}
\affiliation{The Rudolf Peierls Centre for Theoretical Physics, Oxford University, Oxford, OX1 3NP, United Kingdom}
\author{Jacopo De Nardis}
\affiliation{D\'epartement de Physique, \'Ecole Normale Sup\'erieure / PSL Research
University, CNRS, 24 rue Lhomond, 75005 Paris, France}
\author{Maurizio Fagotti}
\affiliation{D\'epartement de Physique, \'Ecole Normale Sup\'erieure / PSL Research
University, CNRS, 24 rue Lhomond, 75005 Paris, France}
\begin{abstract}
We consider the non-equilibrium time evolution of piecewise homogeneous states in the XXZ \mbox{spin-${1}/{2}$} chain, a paradigmatic example of an interacting integrable model. 
The initial state can be thought as the result of joining chains  with different global properties. Through dephasing, 
at late times the state becomes locally equivalent  to a stationary state which explicitly depends on position and time.    
We propose a kinetic theory of elementary excitations and derive a continuity equation which fully characterizes the thermodynamics of the model. We restrict ourselves to the gapless phase and consider cases where the chains are prepared: 1) at different temperatures; 2) in the ground state of two different models; 3) in the ``domain wall'' state.  
We find excellent agreement (any discrepancy is within the numerical error) between theoretical predictions and numerical simulations of time evolution based on \textsc{tebd} algorithms. As a corollary, we unveil an exact expression for the expectation values of the charge currents  in a generic stationary state. 
\end{abstract}
\maketitle

During the last decade, the study of non-equilibrium dynamics in quantum many-body systems has experienced a golden age. The experimental possibility for investigating almost purely unitary time evolution~\cite{exp} sparked off a diffuse theoretical excitement~\cite{EF:review, P:review, quench,GE15,dephasing, IlievskiPRLQL}. The challenge was to understand in which sense unitarily evolving systems can relax to stationary states, and, if this happens, how to determine the stationary values of the observables. The main focus has been on translationally invariant systems. There, a clear theoretical construction has been developed:  while the full system can not relax, in the thermodynamic limit finite subsystems can, as the rest of the system acts as an unusual bath. It was argued that the stationary values of local observables are determined by local  and quasi-local conservation laws~\cite{Doyon,P:review, EF:review}.  It is then convenient to distinguish between \emph{generic models}, where the Hamiltonian is the only local conserved quantity, and \emph{integrable models}, where the number of local charges scales with the systems's size. It was conjectured that in the former case stationary values of local observables are described by Gibbs ensembles (\textsc{ge}) \cite{thermalization} while in the latter by so-called generalised Gibbs ensembles (\textsc{gge}) \cite{Rigol07}. 
Importantly, traces of the underlying integrability remain even in the presence of small integrability-breaking perturbations: at intermediate times the expectation values of local observables approach quasi-stationary plateaux retaining infinite  memory of the initial state~\cite{prethermalization, prerelaxation, LGS:review, prethermalizationexp}.

In the absence of translational invariance the situation gets more complicated. In this context a variety of different settings have been considered, which can be cast into two main classes. The first consists of dynamics governed by translationally invariant Hamiltonians on inhomogeneous states. Relevant examples are the sudden junction of two chains at different temperature~\cite{thermoelectricengine, BD:review, VM:review, twotemperatures, conjTT, Zotos, CAD:hydro,2Delta}, with different magnetizations~\cite{Antal, twomagnetizations}, or with other different global properties~\cite{otherdifferentglobalproperties,XXlqss}. 
In the second class we include dynamics where the Hamiltonian features a localised defect~\cite{Fdefect, BF:NESSdef,SM:toc, CVZ:14}. In both cases, a non-equilibrium steady state (\textsc{ness}) emerges: around the junction of the chains in the first class of problems~\cite{Ruelle,BD:nessCFT, CAD:hydro} and close to the defect in the second~\cite{Fdefect, BF:NESSdef}. 
The characterization of the 
transport properties of the \textsc{ness} have attracted tremendous attention; however, the \textsc{ness} is just the tip of the iceberg.
In the limit of large time $t$ and large distance $x$ from the inhomogeneity, the state becomes locally equivalent to a nontrivial stationary state, which, in integrable models, turns out to depend only on the ``ray'' $\xt=x/t$~\cite{Antal,BF:NESSdef,XXlqss}. We will refer to the latter as a locally-quasi-stationary state (\textsc{lqss})~\cite{BF:NESSdef}. 
We note that ray-dependent profiles of specific observables emerge naturally in hydrodynamical approaches \cite{A:hydro}, which have been applied also in 
more generic systems \cite{profilesgenericmodels}.
Even though these problems have been under scrutiny for a long time,  exact analytic results have been obtained only in noninteracting models and conformal field theories, the role of  interaction remaining elusive until now.

In this Letter we study transport phenomena in interacting integrable models, focusing on the first class of protocols. We propose a ``kinetic theory'' of the elementary excitations and obtain a continuity equation whose solution gives the \emph{exact} \textsc{lqss} characterizing the state of the system at late times. 
Solving the continuity equation gives us full access to the state and, in particular, to the expectation values of charge densities and related currents in the entire light cone. To illustrate our ideas, we use the paradigmatic example of the XXZ model.

\paragraph*{The model.} %

We consider the XXZ spin-$1/2$ chain described by the Hamiltonian 
\be
{\boldsymbol H}=J\sum\nolimits_{\ell=1}^{L}\Bigl[{\boldsymbol s}^{x}_\ell {\boldsymbol s}^{x}_{\ell+1}+  {\boldsymbol s}^{y}_\ell {\boldsymbol s}^{y}_{\ell+1}+\Delta {\boldsymbol s}^{z}_\ell {\boldsymbol s}^{z}_{\ell +1}\Bigr],
\label{Eq:Hamiltonian}
\ee
where $L$ is the chain's length, bold symbols indicate quantum operators, and $\{{\boldsymbol s}^{\alpha}_\ell\}$ are spins ${1}/{2}$. 
We consider $|\Delta|\leq 1$, parametrise the anisotropy as $\Delta=\cos(\gamma)$ and set $J=1$. The model is solved by the Bethe ansatz~\cite{Bethe}: every eigenstate $\ket{\{\lambda_i\}}$ is parametrised by a set of $N$ complex ``rapidities''  $\{\lambda_i\}$ fulfilling the Bethe equations 
\be
\Bigl[\frac{\sinh\left(\lambda_j+i\frac{\gamma}{2}\right)}{\sinh\left(\lambda_j-i\frac{\gamma}{2}\right)}\Bigr]^{L}
=\prod\nolimits_{\substack{l\neq j}}^N\Bigl[\frac{\sinh\left(\lambda_j-\lambda_l+i\gamma\right)}{\sinh\left(\lambda_j-\lambda_l-i\gamma\right)}\Bigr].
\label{Eq:Bethe}
\ee
Following the ``string hypothesis'' \cite{Takahashibook}, as $L\rightarrow\infty$ the solutions to \eqref{Eq:Bethe} are organised in  different types of ``string'' patterns, composed by a set of rapidities with the same real part and equidistant imaginary parts. The different string types are interpreted as different species of quasi-particles with real rapidites. In the thermodynamic limit $L\rightarrow\infty$ with $N/L$ fixed, the thermodynamic Bethe ansatz formalism (\textsc{tba}) applies; a thermodynamic state is parametrised by 
``particles'' and ``holes'' distributions $\{\rho_k, \rho^h_{k}\}$, one for each species of quasi-particles. These distributions, usually called ``root densities'', are connected to one another through the thermodynamic version of \eqref{Eq:Bethe}, reported in \cite{SM}. 
The number of species is finite when $\gamma$ is a rational multiple of $\pi$, which is the case considered in this paper. The expectation value of the density $\boldsymbol{q}$ of a conserved charge $\boldsymbol{Q}$ in the stationary state $\ket{\rho}$ reads as~\cite{Korepinbook}
\be\label{eq:charge}
\braket{\rho|\boldsymbol{q}|\rho}=\sum\nolimits_{k}\int{\rm d}\mu\, q_k(\mu) \rho_k(\mu)\,,
\ee
where $q_k(\mu)$ is the  single-particle eigenvalue of the charge and is independent of the state. 
If $\ket{\rho}$ is invariant under spin-flip $\prod_j\sigma_j^x$,  it is completely characterized by the expectation values of the local and quasi-local charges obtained from the unitary representations of the transfer matrix
~\cite{IlievskiPRLQL, P:review}. We indicate these charges by $\boldsymbol{Q}_n^{(s)}$, with $n,2s\in \mathbb N$, and the single-particle eigenvalues by $q^{(s)}_{n,k}(\mu)$. The charges have an increasing typical range as a function of $n$ and
$q_{n,k}^{(s)}(\mu)=-({\sin\gamma}/{2})\partial_\mu q^{(s)}_{n-1,k}(\mu)$; in particular, ${\boldsymbol Q}_n^{(1/2)}$ are local~\cite{Korepinbook} and ${\boldsymbol Q}_1^{(1/2)}=\boldsymbol{H}-\frac{\Delta L}{4}$. 
We refer the reader to the Supplemental Material~\cite{SM} and to the specific literature~\cite{Takahashibook,Korepinbook,P:review} for further details. 
A case without spin-flip invariance is discussed in {\tt Example\,3}.

\paragraph*{Locally-quasi-stationary state.} 
In integrable models, the information about an inhomogeneity spreads linearly in time because of stable quasi-particle excitations~\cite{bonnes14}. 
These contribute to the emergence of non-trivial behavior   
along the rays $\xt=x/t$. Dephasing mechanisms~\cite{dephasing} 
are active also in the inhomogeneous case so at sufficiently late times the dynamics are expected to slow down with an emergent timescale proportional to~$x$.
 Thus we assume that, for given $\xt$, the expectation values of observables can be eventually described by a stationary state $\boldsymbol \rho^{\textsc{lqss}}_\xt$
\be
\braket{\boldsymbol{\mathcal O}}_{x,t}\equiv\braket{\Psi_t|{\boldsymbol{\mathcal O}_x}|\Psi_t}=\tr{}{\boldsymbol \rho^{\textsc{lqss}}_\xt\boldsymbol{\mathcal O}_x}+o(t^{-\epsilon})\, .
\ee
Here $\boldsymbol{\mathcal O}_{x}$ acts non-trivially only around $x$. 
The state $\boldsymbol \rho^{\textsc{lqss}}_{\xt}$ is the \textsc{lqss} introduced in~\cite{BF:NESSdef}; determining it exactly is our main goal. 

\paragraph*{Kinetic Theory.}  %

Being stationary (for given $\xt$), $\boldsymbol \rho^{\textsc{lqss}}_\xt$ is characterized by a set of root densities $\{ \rho_{\xt, j},  \rho^h_{\xt, j}\}$: 
\be
\tr{}{{\boldsymbol \rho}^{\textsc{lqss}}_{\xt}\boldsymbol{\mathcal{O}} }=\braket{\rho_\xt|\boldsymbol{\mathcal{O}}|\rho_\xt}\,.
\ee
In particular, the charges can be written as in \eqref{eq:charge}.

Since the root densities are fixed by the expectation values of the charges~\cite{IlievskiPRLQL}, the full \textsc{lqss} can be obtained by determining how their expectation values vary in time. 
We assume that the change is induced by the motion of elementary excitations and that the late time regime is characterized by a ``dynamical equilibrium'', where the thermodynamic state varies
only slightly even though a \emph{macroscopic} number of quasi-particles is moving. The nature of quasi-particle excitations remains well defined while moving through the system; on the other hand, the excitation energy $\varepsilon_{\xt,k}(\lambda)$ and the momentum $p_{\xt,k}(\lambda)$ depend on the macro-state~\cite{bonnes14}, so 
 the ``mild'' inhomogeneity of the \textsc{lqss}  modifies the propagation velocity $
v_{\xt,k}(\lambda)= {\partial_\lambda \varepsilon_{\xt,k}(\lambda)}/{\partial_\lambda p_{\xt,k}(\lambda)}
$. This leads to
\be
\braket{\boldsymbol{q}}_{x,t + \delta t}-\braket{\boldsymbol{q}}_{x,t}=\int\!{\rm d}\tilde x\,\bigl(\underset{\tilde x\rightarrow  x,t}{\Delta^{\boldsymbol{q}}}- \underset{x\rightarrow  \tilde x, t}{\Delta^{\boldsymbol{q}}}\bigr)\, ,
\label{Eq:dtq}
\ee
where $\underset{\tilde x\rightarrow   x, t}{\Delta^{\boldsymbol{q}}}$ is the charge density $\boldsymbol{q}$ carried from $\tilde x$ to $x$ by the quasi-particles in the time interval $[t,t+\delta t]$. 
For given  $\tilde x-x$ and $\delta t$, only excitations with velocity $v=(x-\tilde x)/\delta t$ contribute to $\underset{\tilde x\rightarrow   x, t}{\Delta^{\boldsymbol{q}}}$, namely
\be
\underset{ \tilde x\rightarrow   x, t}{\Delta^{\boldsymbol{q}}} \equiv \sum\nolimits_{k}\int\!\! {{\rm d}\lambda}\, \delta({ x-\tilde x-v_{\tilde\xt, k}(\lambda)\delta t}) c_{k}^{\boldsymbol{q}}(\lambda|\tilde\xt).
\label{Eq:kinetic}
\ee    
Here $c_{k}^{\boldsymbol{q}}(\lambda|\xt)\mathrm d\lambda$ is the charge density transported by excitations with string type $k$ and rapidity $\in[\lambda,\lambda+d\lambda]$. This quantity depends on $\xt$ through ${\boldsymbol \rho}^{\textsc{lqss}}_{\xt}$ and will be expressed in terms of the root densities $\rho_{\xt,j}$ before long. Plugging \eqref{Eq:kinetic} into \eqref{Eq:dtq} gives
\be
\partial_t\braket{\boldsymbol{q}}_{x,t }=-\sum\nolimits_{k}\int\! {{\rm d}\lambda}\,\partial_x\left[v_{\xt, k}(\lambda) c_{k}^{\boldsymbol{q}}(\lambda|\xt) \right]\,.
\ee 
By virtue of \eqref{eq:charge}  we then find 
\begin{align}
\!\!\!\sum\nolimits_{k}\!\!\int\!\! {{\rm d}\lambda}\,\bigl[q_{k}(\lambda)\partial_t \rho_{\xt, k}(\lambda)+\partial_x\!\!\left(v_{\xt,k}(\lambda) c_{k}^{\boldsymbol{q}}(\lambda|\xt)\right)\bigr]=0\,.
\label{Eq:contcharge}
\end{align}
The next step is to fix the form of $c_{k}^{\boldsymbol{q}}(\lambda|\xt)$ in terms of the root densities.
To this aim, it is convenient to consider an auxiliary toy problem as follows. 
Let a macroscopic subsystem $A$ be described by  $\ket{\rho}_A$ with all the root densities equal to zero except for $\rho_k(\lambda)$, with $\lambda\in[\bar\lambda,\bar\lambda+\epsilon]$ and $\epsilon$ some small parameter. Let us then release the subsystem in the vacuum ($\rho_j(\lambda)=0$), namely in an infinite bath of spins  up $\ket{\Psi_0}=\ket{\rho}_A\otimes \ket {\uparrow\cdots\uparrow}_B$.
After a sufficiently long time it is reasonable to expect local relaxation
to the vacuum. From \eqref{eq:charge} it follows that the total charge  density $\Delta q$ flowed out of the subsystem reads $\Delta q =\int_{\bar\lambda}^{\bar\lambda+\epsilon} {\rm d}\lambda\, q_k(\lambda)\rho_k(\lambda)$. Crucially, we interpret this expression as the charge density  $ c_{k}^{\boldsymbol{q}}(\lambda)\epsilon$ associated with the quasi-particles of species $k$ and rapidity $\lambda\in[\bar\lambda,\bar\lambda+\epsilon]$  going out of the subsystem~\cite{f:dr}
\be
c_{k}^{\boldsymbol{q}}(\lambda)=q_k(\lambda)\rho_k(\lambda)\,.
\label{Eq:hyp}
\ee
 Let us go back to the expression \eqref{Eq:contcharge} and take \eqref{Eq:hyp} as the transported charge density; we find
\begin{align}
\!\!\sum\nolimits_{k}\!\!\int\!\! {{\rm d}\lambda}\,q_{k}(\lambda)\bigl[\partial_t \rho_{\xt, k}(\lambda)+\partial_x\left(v_{\xt,k}(\lambda)\rho_{\xt,k}(\lambda)\right)\bigr]\!\!=0\,.
\label{Eq:contchargefin}
\end{align}
Since $q_{k}(\lambda)$ is independent of $\xt$, \eqref{Eq:contchargefin} is a continuity equation for the charge density and holds for any local and quasi-local charge $\boldsymbol{Q}^{(s)}_{n}$. Using the completeness of the set $\{q^{(s)}_{n,k}(\lambda)\}$ we have
\begin{empheq}[box=\fbox]{equation}
\partial_t \rho_{\xt, k}(\lambda)+\partial_x\left(v_{\xt,k}(\lambda)\rho_{\xt, k}(\lambda)\right)=0
\label{Eq:continuity}
  \end{empheq}
This is our main result: the root densities $\rho_{\xt, k}(\lambda)$, characterizing the state, obey a continuity equation with a $\xt$-dependent velocity, remarkable effect of the interaction that induces a state-dependent dressing on the elementary excitations. 
\emph{A priori}, one would expect the physical picture based on a kinetic theory of excitations to be only approximately correct. 
In fact, we will provide evidence that \eqref{Eq:continuity} \emph{exactly}  describes the dynamics at late times $t$ and large distances $x$ along the ray $\xt=x/t$. 

\paragraph*{Charge currents.} %

\begin{figure}[t]
\includegraphics[width=0.45\textwidth]{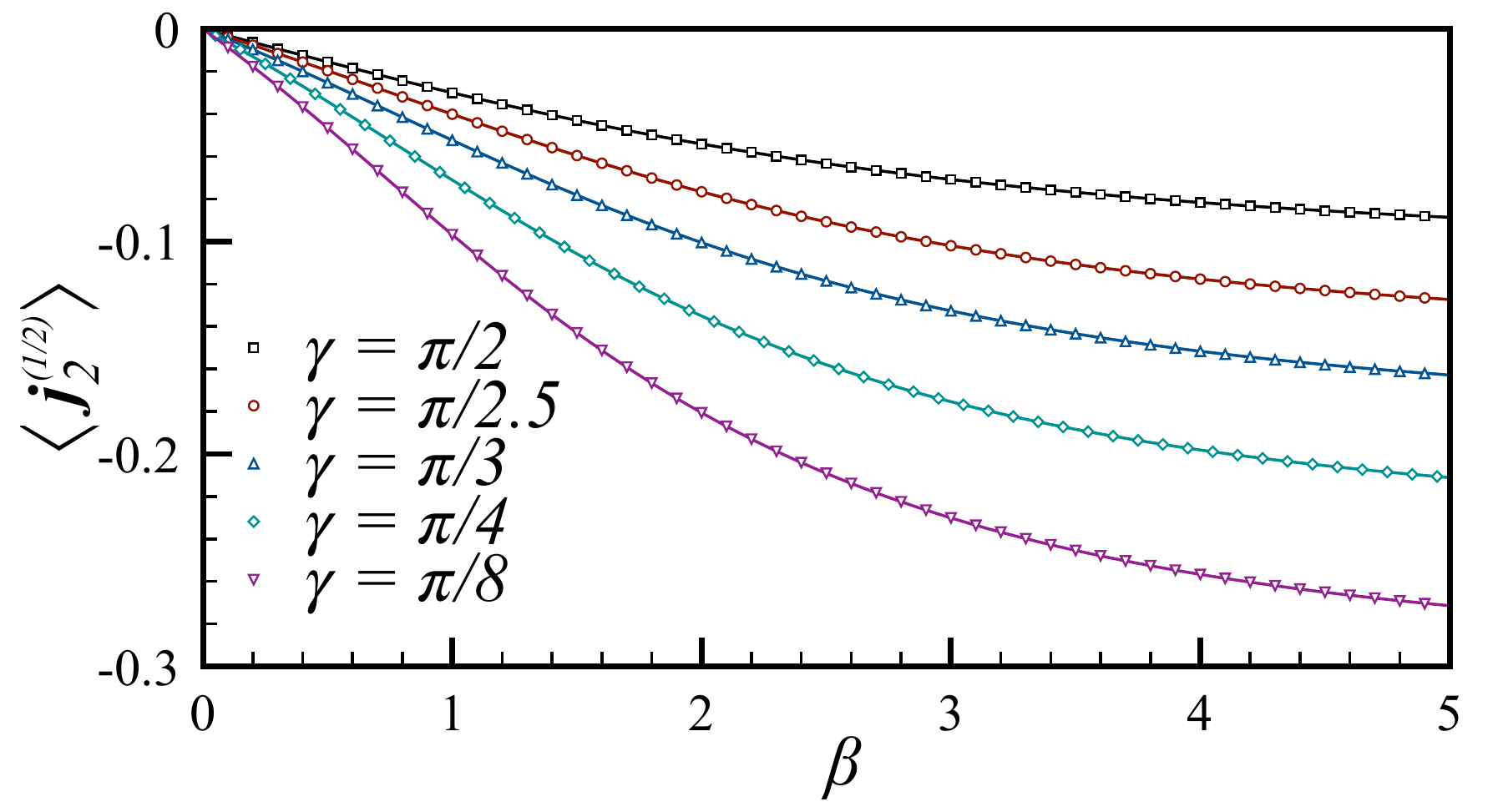}\vspace{-0.25cm}
\caption{Thermal expectation value of  ${\boldsymbol j}^{(1/2)}_{2}$ for a wide range of temperatures and
different anisotropies $\Delta = \cos(\gamma)$. 
Full lines are \textsc{mpdo} data (error $<10^{-6}$) in a system of length $L=50$. Symbols are the prediction (\ref{Eq:current}). }
\label{fig:J2_thermal}
\end{figure}

In a spin chain the current  ${\boldsymbol j}_{\ell}[\boldsymbol{Q}]$ of a charge $\boldsymbol{Q}=\sum_\ell  {\boldsymbol q}_{\ell}$ is defined through the following continuity equation 
\be
{\boldsymbol j}_{\ell+1}[\boldsymbol{Q}]-{\boldsymbol j}_{\ell}[\boldsymbol{Q}]=i [{\boldsymbol q}_{\ell},\boldsymbol{H}]\,.
\ee
Imposing $\tr{}{{\boldsymbol j}_{\ell}[\boldsymbol{Q}]}=0$, this determines ${\boldsymbol j}_{\ell}[\boldsymbol{Q}]$ up to operators with zero expectation value in any translationally invariant state. 
In the infinite time limit along the ray $\xt={x}/{t}$ the time-evolving state becomes homogeneous, so the expectation values of the currents are  independent of their particular definitions.  
From \eqref{Eq:contchargefin} it follows 
\be
\braket{\rho|\boldsymbol{j}_{\ell}[\boldsymbol{Q}]|\rho} \sim \sum\nolimits_k\int\mathrm d \lambda  q_k(\lambda)v_k(\lambda)\rho_k(\lambda)
\label{Eq:current}
\ee 
where  $\ket{\rho}$ is an arbitrary stationary state and the equivalence is up to a state-independent constant.  

We now provide several compelling consistency checks for the validity of \eqref{Eq:current} and, in turn, of \eqref{Eq:contchargefin}.

{\rm\underline{\tt Check\,1}\label{check1}: \emph{Conservation of the energy current.}}
In the XXZ model the energy current is equal to the second charge, namely: ${\boldsymbol j}^{(1/2)}_{1,\ell}\sim {\boldsymbol q}_{2,\ell}^{(1/2)}$, where we introduced  the notation ${\boldsymbol j}^{(s)}_{n,\ell}\equiv {\boldsymbol j}_{\ell}[\boldsymbol{Q}^{(s)}_n]$. 
Using some \textsc{tba} identities one can easily show that this relation is satisfied by \eqref{Eq:current}~\cite{SM}.

{\rm\underline{\tt Check\,2}: \emph{Current(s) at equilibrium vs numerics}.}
Fig.~\ref{fig:J2_thermal} shows the  expectation value of the current $\boldsymbol{j}_{2,\ell}^{(1/2)}$ in thermal states with inverse temperature $\beta\in[0,5]$ and for different values of $\Delta$. 
The prediction \eqref{Eq:current} is checked against numerical data obtained using an algorithm based on the Matrix Product Density Operator (\textsc{mpdo}) representation of a mixed state~\cite{SM}. The agreement is unquestionably perfect: the discrepancies are smaller than the \textsc{mpdo} accuracy.

{\rm\underline{\tt Check\,3}: \emph{Comparison with other results}.}
Ref.~\cite{CAD:hydro} independently obtained an expression for the currents in an integrable quantum field theories with diagonal scattering. In \cite{SM} it is shown that this is equivalent to \eqref{Eq:current}. 

\paragraph*{Determining the \textsc{lqss}.} %

\begin{figure}[t]
\includegraphics[width=0.225\textwidth]{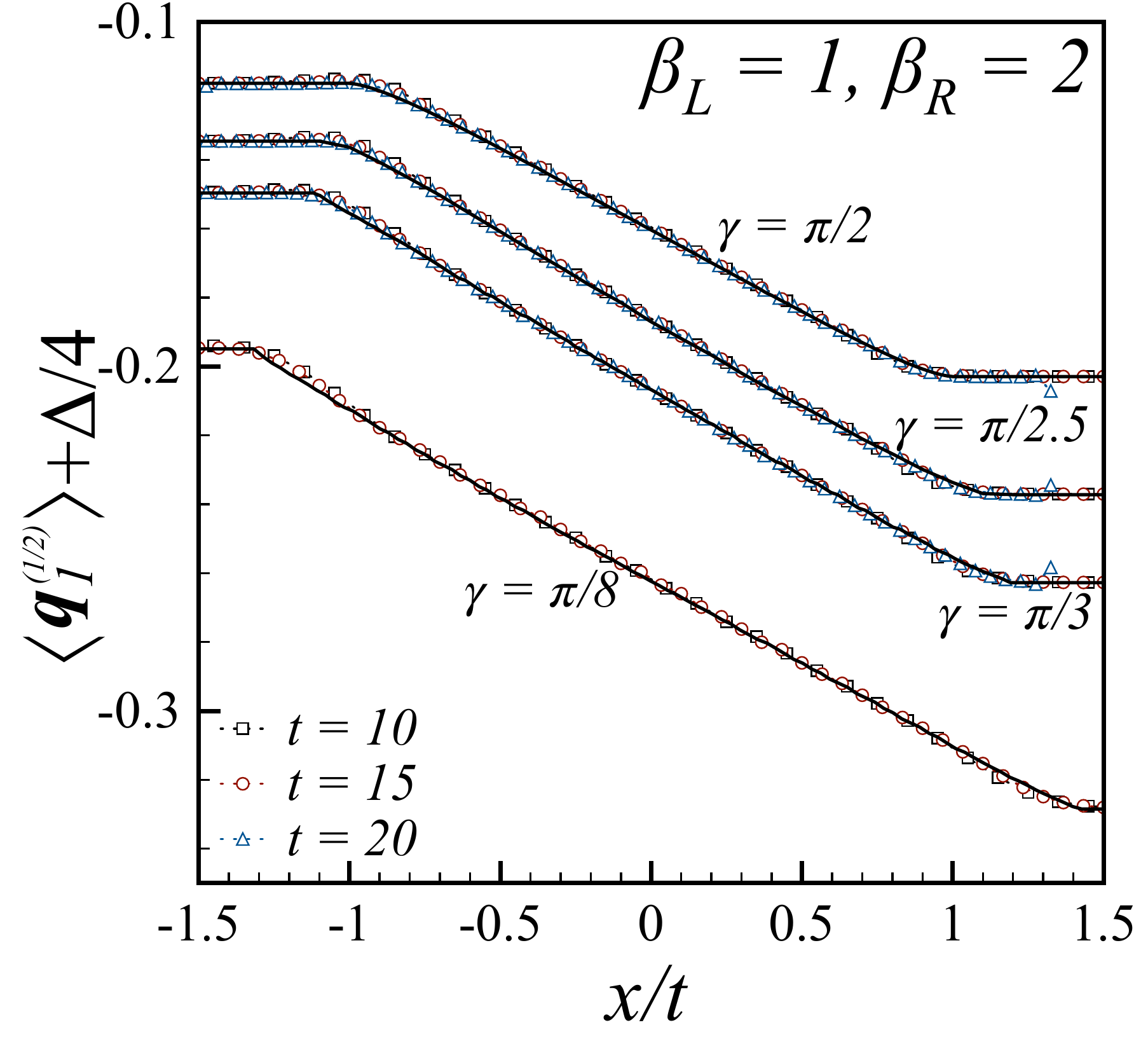}\includegraphics[width=0.225\textwidth]{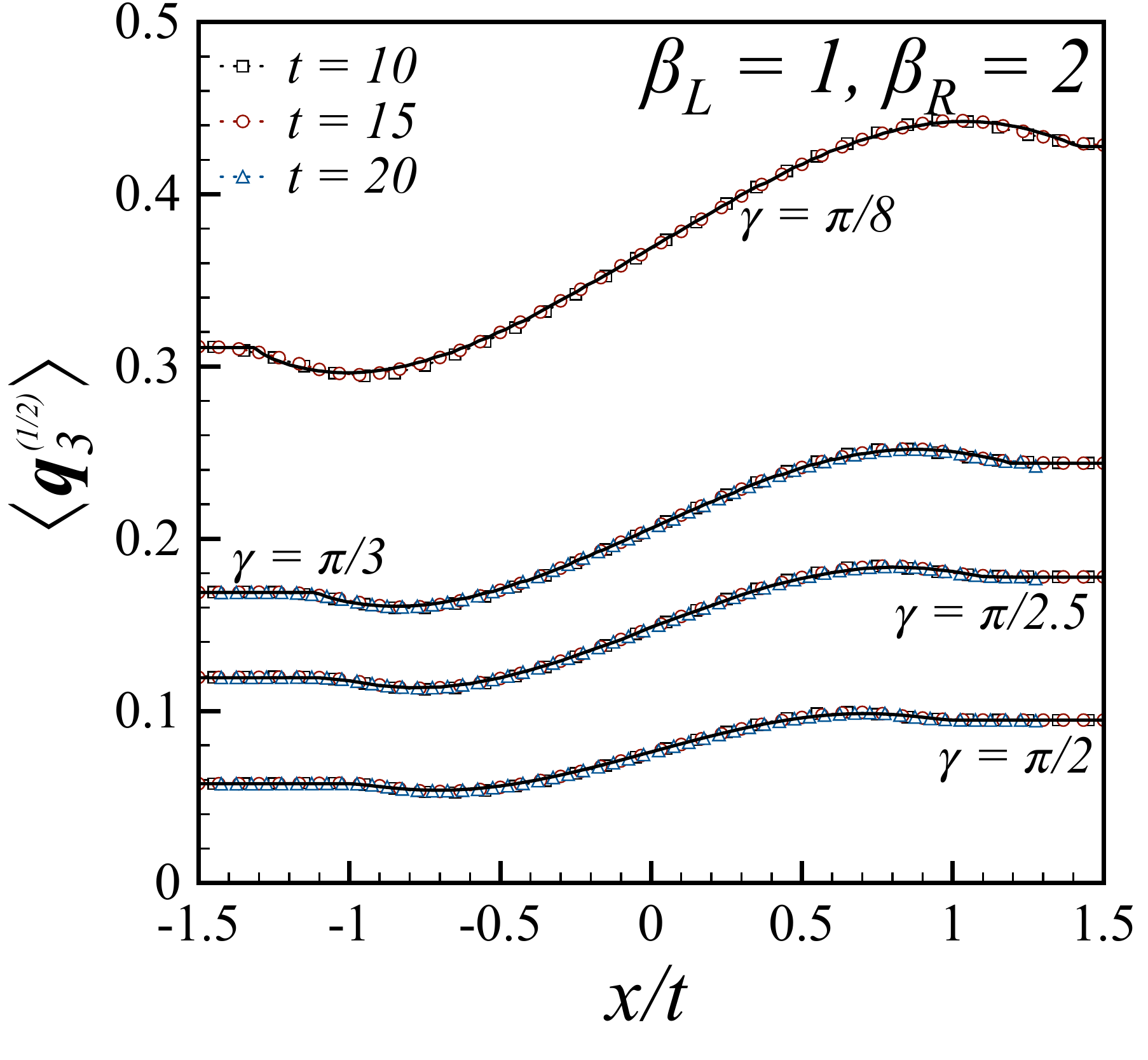}\\
\includegraphics[width=0.225\textwidth]{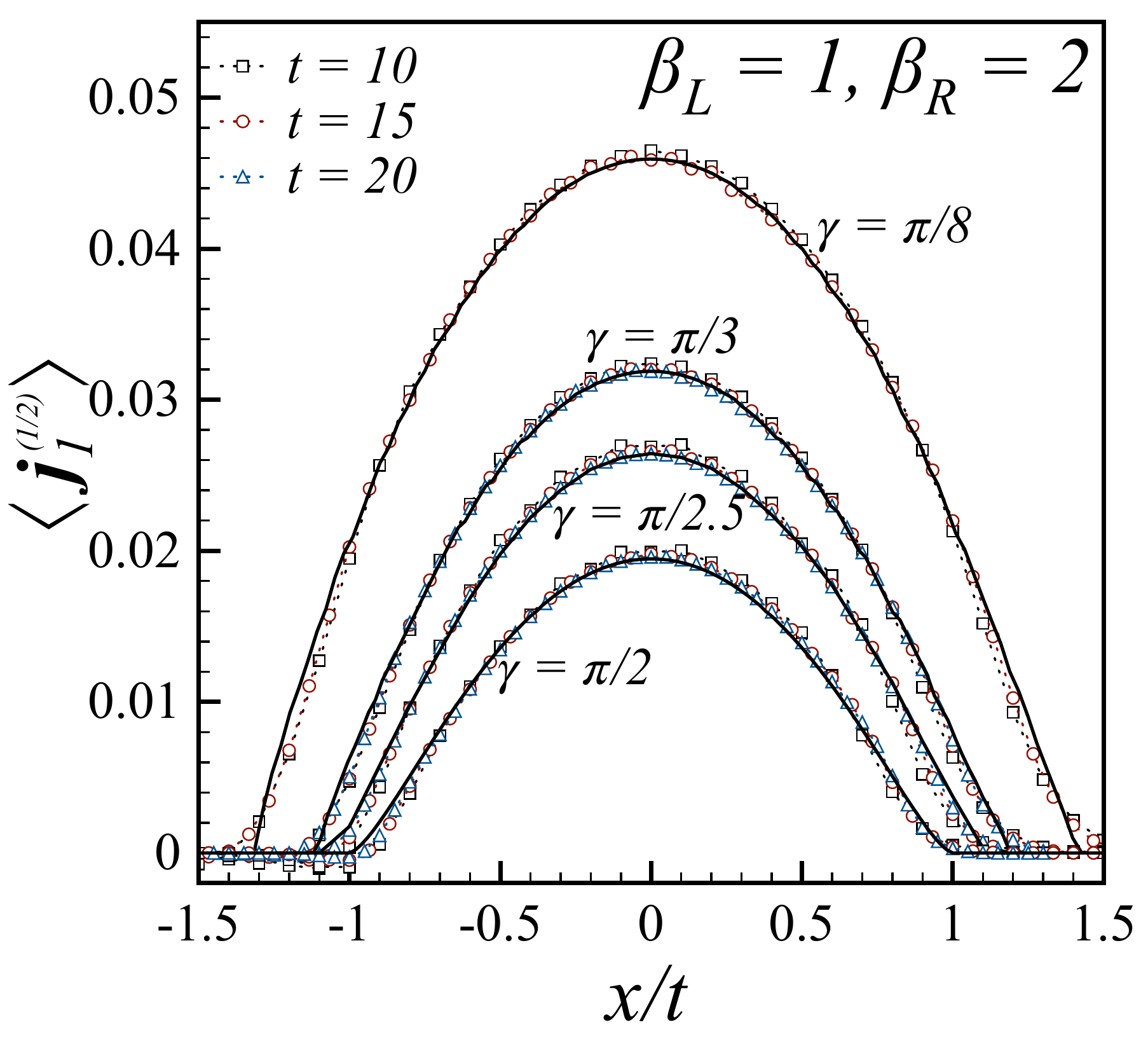}\includegraphics[width=0.225\textwidth]{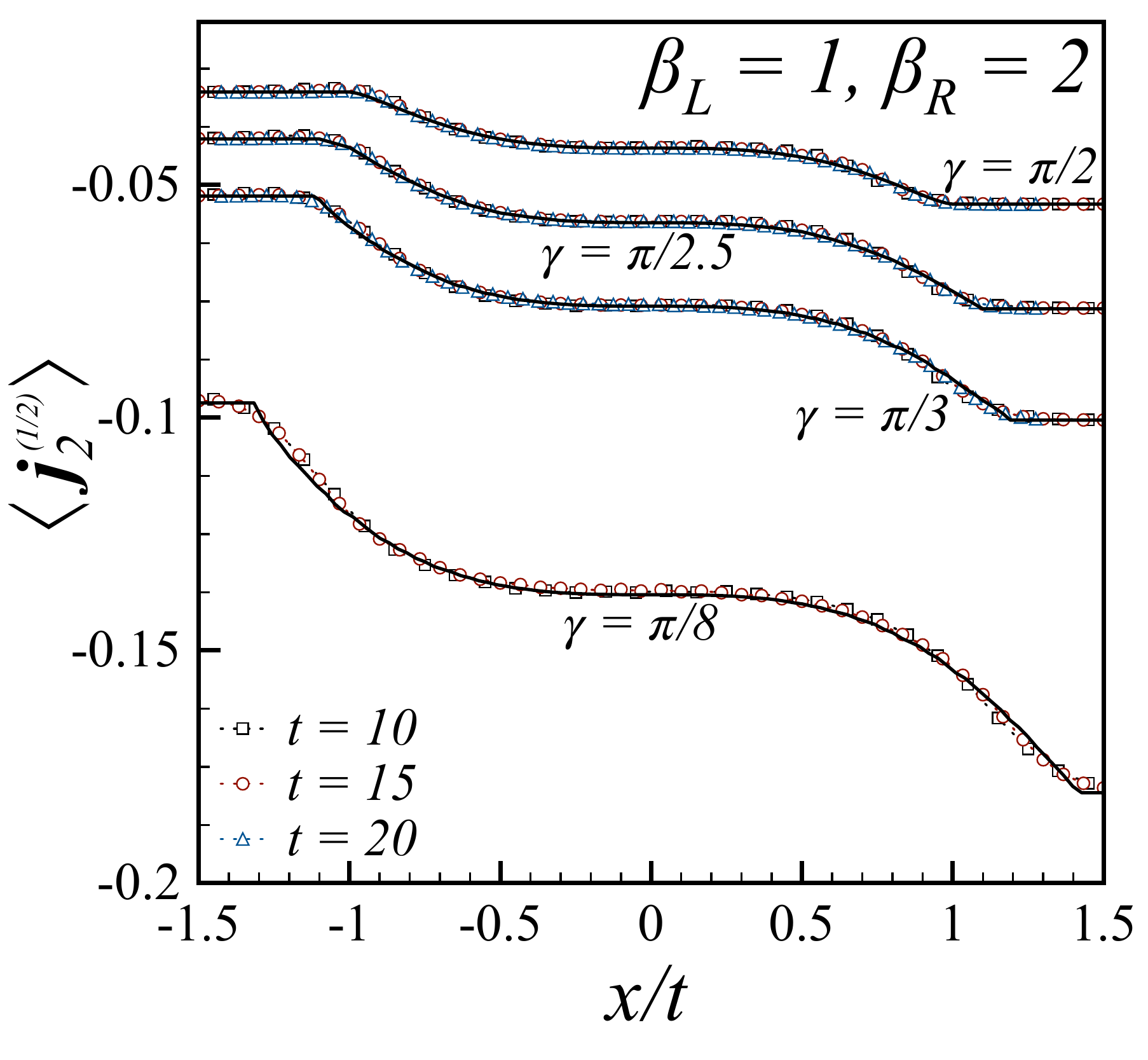}
\vspace{-0.25cm}
\caption{Profiles of charge densities and currents 
for three different values of $\Delta = \cos(\gamma)$.  A 60-sites chain has been initially prepared in two halves at inverse temperatures 
$\beta_{L}=1$ and $\beta_{R}=2$. Symbols denote \textsc{mpdo} data; full black lines are the predictions based on \eqref{eq:solution}.  
The tiny ripples in the predictions are numerical artifacts.}
\label{fig:betaL1_betaR2}
\end{figure}

\begin{figure}[t!]
\includegraphics[width=0.45\textwidth]{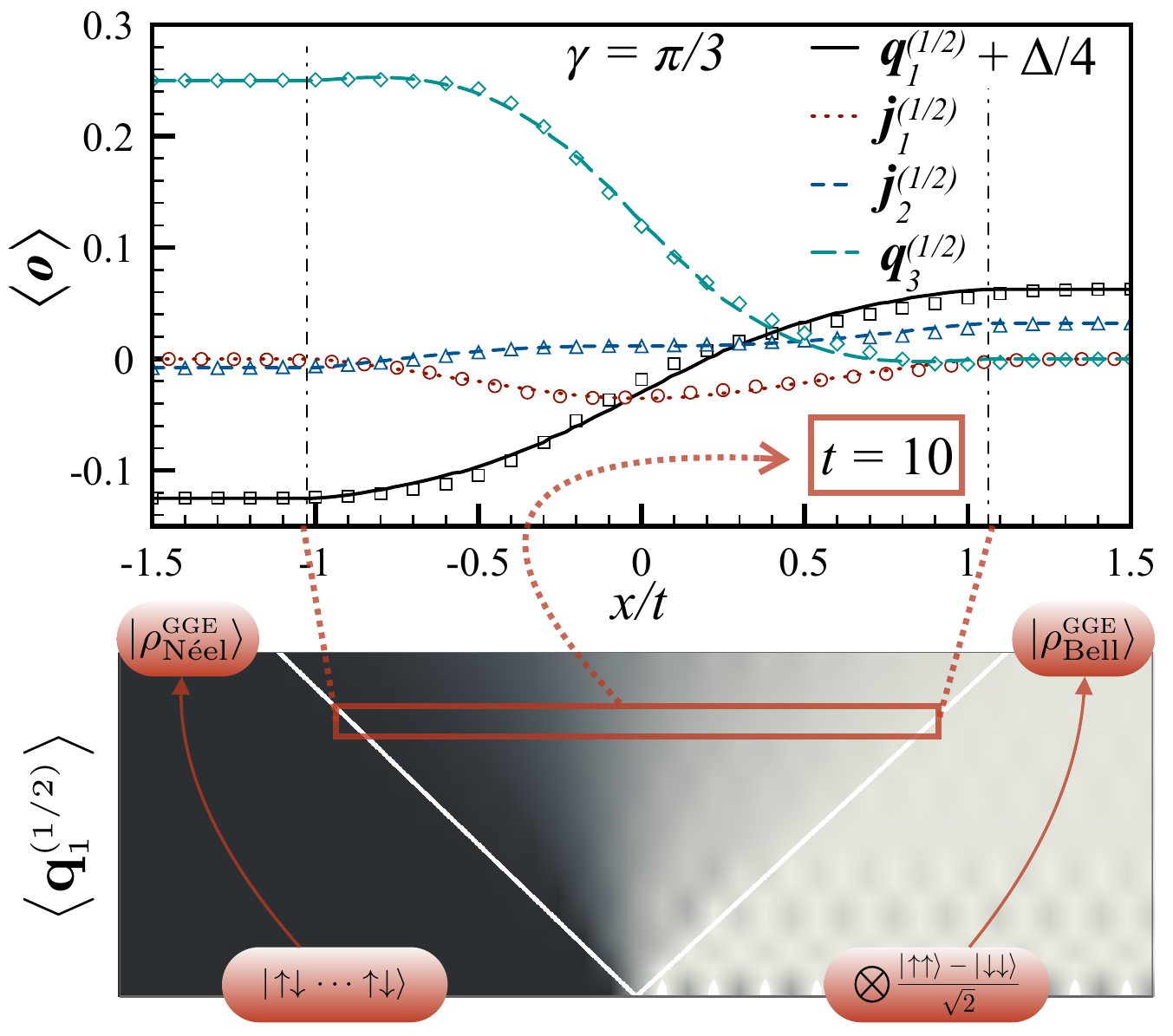}
\caption{
\textsc{up}:
Profiles of different charge densities and currents for the quench of {\tt Example\,2} with $\Delta = \cos(\pi/3)$.
Predictions of Eq.~\eqref{eq:solution}
(lines) are compared with \textsc{tebd} data at time $t=10$ (symbols) obtained in a 100-sites chain. 
Spatial oscillations in the \textsc{tebd} data  were smoothed out by taking a local spacial average.
The vertical dotted-dashed lines represent the light-cone edges. 
\textsc{down}: Space-time density plot of $\braket{\boldsymbol{q}^{(1/2)}_1}$.}
\label{fig:Delta05_Neel_BELL}
\end{figure}

We now turn to our main goal: the determination of the \textsc{lqss} evolving from an inhomogeneous state. We consider the time evolution of $\ket{\psi_0}^{\textsc{l}}\otimes\ket{\psi_0}^{\textsc{r}}$  under the Hamiltonian \eqref{Eq:Hamiltonian} at sufficiently long times. 
The dynamics is described by \eqref{Eq:continuity}, which, using some \textsc{tba} identities~\cite{SM}, can be recast in the form
\be\label{eq:cont}
[\xt- v_{\xt, k}(\lambda)]\partial_\xt\vartheta_{\xt,k}(\lambda)\rho_{\xt,k}^{\, t}(\lambda)=0\, ,
\ee
where $\vartheta_{\xt,k}(\lambda)\equiv\rho_{\xt,k}(\lambda)/[\rho_{\xt,k}(\lambda)+\rho^h_{\xt,k}(\lambda)]$. Since $\rho_{\xt,k}(\lambda)+\rho^h_{\xt,k}(\lambda)>0$,  the solution $\vartheta_{\xt, k}(\lambda)$ is a piecewise constant function of $\xt$. If, for any $\lambda$, $v_{\xt,k}(\lambda)=\xt$ has a unique solution~\cite{note1} we find
\be\label{eq:solution}
\vartheta_{\xt,k}(\lambda)=\theta_{\textsc{h}}(v_{\xt,k}(\lambda)-\xt)(\vartheta^{\textsc{l}}_k(\lambda)-\vartheta^{\textsc{r}}_k(\lambda))+\vartheta^{\textsc{r}}_k(\lambda)\, .
\ee
Here $\theta_{\textsc{h}}(x)$ is the step function which is nonzero and equal to $1$ only if $x>0$. 
The functions  $\vartheta^{\textsc{l}}_k(\lambda)$ and $\vartheta^{\textsc{r}}_k(\lambda)$ are the boundary conditions: due to the Lieb-Robinson bounds~\cite{{LR72},{bravyi06}}, there exists a maximal velocity $v_{\rm max}$ such that observables on rays $|\xt| > v_{\rm max}$ never receive information about the inhomogeneity; as a result, $\vartheta^{\textsc{l}}_k(\lambda)$ and $\vartheta^{\textsc{r}}_k(\lambda)$ describe the stationary states emerging independently in the two, left and right, bulk parts of the system (see also Fig.~\ref{fig:Delta05_Neel_BELL}).

As $v_{\xt,k}(\lambda)$ depends on $\vartheta_{\xt,k}(\lambda)$, \eqref{eq:solution} is only an implicit representation of the solution. In practice, one can solve the problem by iteration, starting from an initial $\vartheta^{(0)}_{\xt, k}(\lambda)$, computing the excitation velocities, and iterating again until convergence is reached. The procedure is numerically very efficient and converges after few iterations. 

{\rm\underline{\tt Example\,1}: \emph{Two temperatures.}}
Let us consider the transport problem \emph{par excellence}: two chains prepared at different temperatures and then joined together~\cite{BD:review}.

In Fig.~\ref{fig:betaL1_betaR2} 
we report the rescaled profiles of a number of charges and currents 
for different times $t=10,\,15,\,20$ and interactions $\Delta$. 
The rescaled numerical data are in excellent agreement with the analytical predictions. This strongly suggests that the solution of \eqref{Eq:continuity} \emph{fully} characterizes the state of the system at late times.

We note that at the edges of the light cone the predictions are not smooth, as the profiles are exactly flat outside the light cone. This is an infinite-time property, and indeed the numerical data are smooth at any time.
Moreover, contrary to the noninteracting case, the velocities also depend on the temperatures~\cite{bonnes14}, as revealed by the slight asymmetry of all the curves reported in Figs~\ref{fig:betaL1_betaR2}.

We mention that the conjecture put forward in~\cite{Zotos} for the energy current $\boldsymbol{j}^{(1/2)}_{1,\ell}$ at $\xt=0$ is only in a fair agreement with our results~\cite{SM}.

{\rm\underline{\tt Example\,2}: \emph{Global quench.}}
We now study the dynamics after joining together two globally different pure states which are not stationary.  
This is a genuine global quench with nontrivial time evolution also outside the light cone. 
As initial state we take the tensor product between the N\'eel state $\ket{\uparrow\downarrow\cdots \uparrow\downarrow}$ and the Bell state $\bigotimes_j\frac{\ket{\uparrow\uparrow}_j-\ket{\downarrow\downarrow}_j}{\sqrt{2}}$. As explained before, the two boundary conditions $\vartheta^{\textsc{l}}, \vartheta^{\textsc{r}}$ are the \textsc{gge}'s corresponding to the quenches $e^{-i \boldsymbol{H} t}\ket{\text{N\'eel}}$ and $e^{-i \boldsymbol{H} t}\ket{\text{Bell}}$ (see the \textsc{down} panel of Fig.~\ref{fig:Delta05_Neel_BELL}).
Relaxation is slower than in the first example and the comparison with the \textsc{tebd} data  shown in Fig.~\ref{fig:Delta05_Neel_BELL} is jeopardised by the smallness of the time reached, consequence of the linear increase of the entropy both inside and outside the light cone. Nevertheless, the agreement is fairly good.

\begin{figure}[t!]
\includegraphics[width=0.225\textwidth]{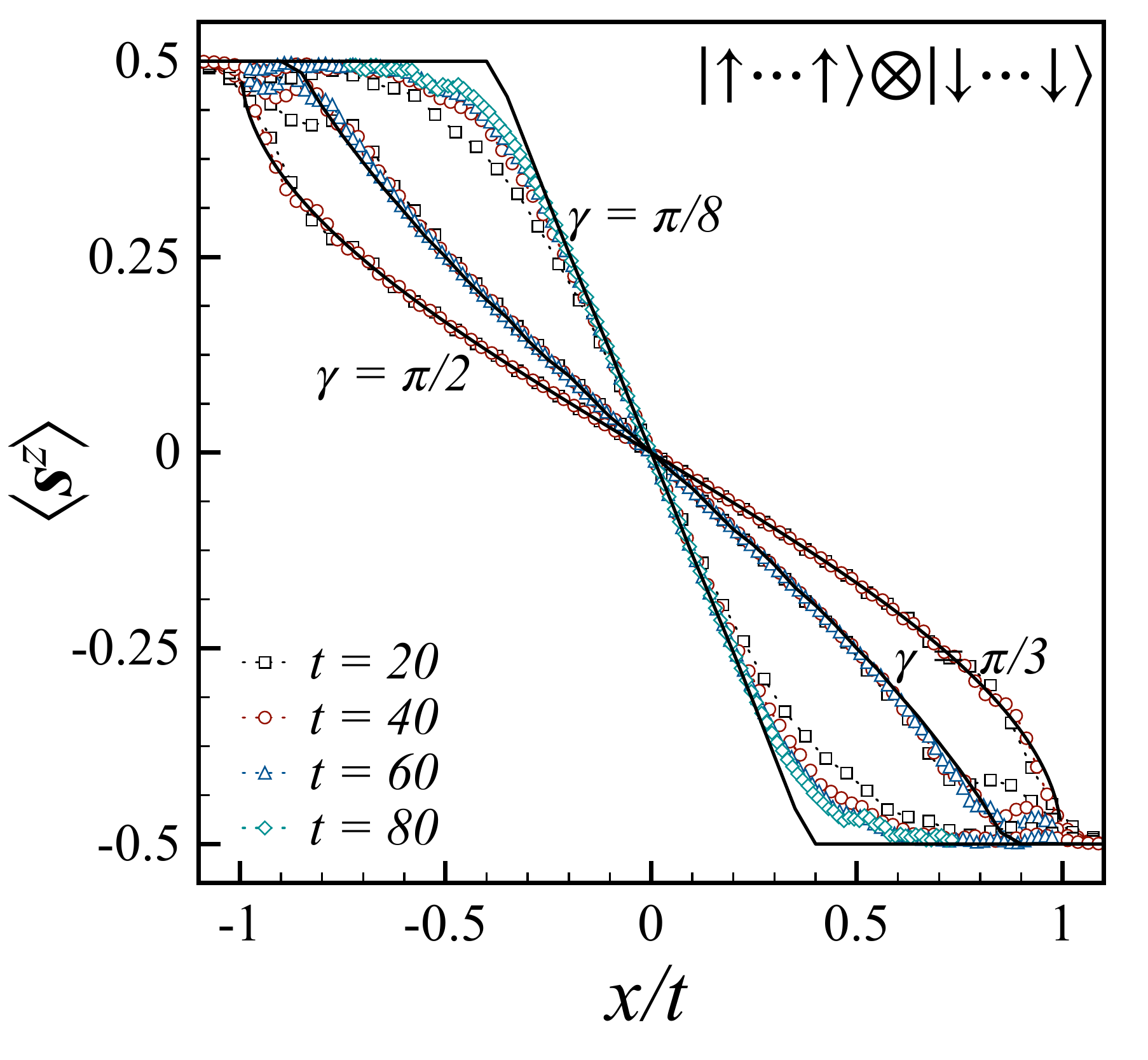}\includegraphics[width=0.225\textwidth]{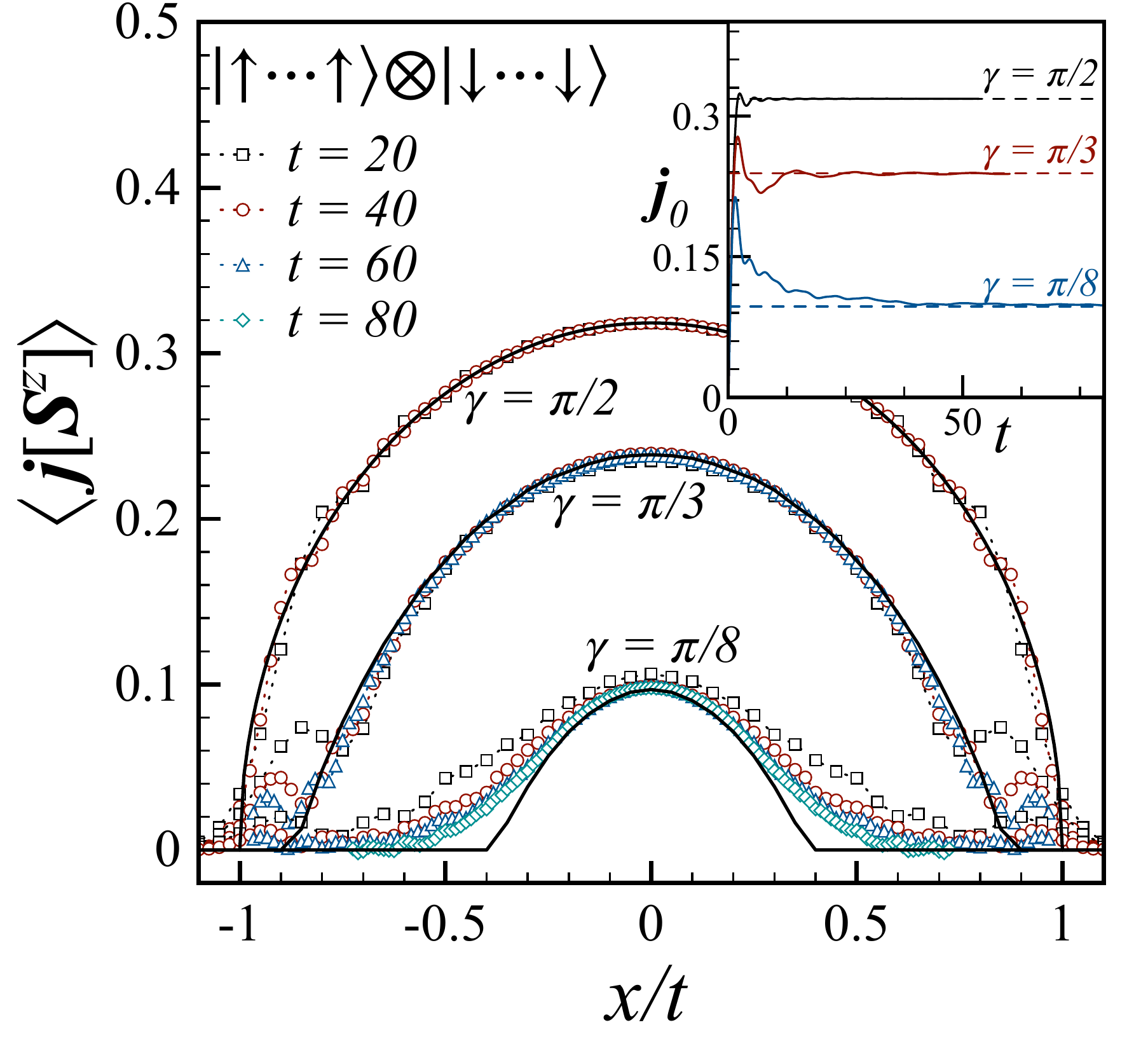}\vspace{-0.25cm}
\caption{Profiles of magnetization 
$\boldsymbol{s}^z_\ell$ and spin-current
 $\boldsymbol{j}_\ell[\boldsymbol{S}^z]$ evolving from a ``domain-wall'' state for three different values of $\Delta = \cos(\gamma)$. Symbols are numerical
data for a 120-sites chain; full black lines are the predictions based on \eqref{eq:solution}. The inset shows the approach of $\boldsymbol{j}_0[\boldsymbol{S}^z]$  (full colored lines) to the prediction (dashed lines).}
\label{fig:DW}
\end{figure}

{\rm\underline{\tt Example\,3}: \emph{Domain wall.}}
If the initial state is not  spin-flip  invariant, the set $\{\boldsymbol{Q}^{(s)}_{n,\ell}\}$ is generally not sufficient to fix the state. First of all, one has to include the total spin along $z$, $\boldsymbol S^z$, but also quasi-local charges coming from non-unitary representations of the transfer matrix~\cite{P:review} might play some role. 
Nonetheless, for a domain-wall initial state~\cite{twomagnetizations} $\ket{\uparrow\ldots\uparrow}\otimes \ket{\downarrow\ldots\downarrow}$, the comparison with numerics provides strong evidence that the expectation values of $\boldsymbol{q}^{(s)}_{n,\ell}$, $\boldsymbol s^z_\ell$, and the corresponding currents, can be obtained from the root densities solving the continuity equation \eqref{Eq:continuity}. The left boundary condition is $\vartheta^{\textsc{l}}_j(\lambda)=0$, while $\vartheta^{\textsc{r}}_j(\lambda)$ corresponds to the state $\propto e^{\mu \boldsymbol S^z}$ in the limit $\mu\rightarrow \infty$. Fig.~\ref{fig:DW} shows the only two measured quantities exhibiting a non trivial behavior. Remarkably, the effective velocities of quasi-particles shrink to zero in the limit $\Delta \to 1$. A more careful analysis will be carried out in a future work.

\paragraph*{Conclusions.} %

Using a ``kinetic theory'' of quasi-particles excitations, we derived a continuity equation (\emph{cf}. \eqref{Eq:continuity}) describing the late time dynamics of the XXZ spin-$1/2$ chain after joining together two macroscopically different homogeneous states. 
We provided compelling evidence that equation \eqref{Eq:continuity} is in actual fact the \emph{exact} continuity equation fulfilled by the conserved charges of the model for late times and large distances.
We tested the predictions for the late-time dynamics against \textsc{tebd} numerical simulations and we have found excellent agreement.  
Our construction is sufficiently generic to be applicable to other interacting integrable models. In addition, the continuity equation can also be applied when integrability is broken by some localized inhomogeneity~\cite{2Delta}: the solution of the late-time dynamics along the rays  originating from the inhomogeneity 
is reduced to the determination of a few boundary conditions.

\begin{acknowledgments} %

We are very grateful to Lorenzo Piroli for the collaboration
at an early stage of this project and for valuable
comments on the manuscript. We thank Vincenzo Alba,
Pasquale Calabrese, Ollala Castro-Alvaredo, Andrea De
Luca, J\'er\^ome Dubail, and Leonardo Mazza for useful
discussions, Benjamin Doyon and Xenophon Zotos, also,
for stimulating correspondence. This work was supported
by the European Research Council under Starting Grant
No. 279391 EDEQS (all authors), by the European Union's
Horizon 2020 research and innovation programme under
the Marie Sklodowska-Curie Grant Agreement No. 701221
(M.C.), and by Laboratory of Excellence ENS-ICFP:ANR-
10-LABX-0010/ANR-10-IDEX-0001-02 PSL* (M.F.,
J.D.N.). J.D.N. and M.F. kindly acknowledge SISSA
for the hospitality during the earlier stage of this work.

Once this paper was being concluded, the preprint \cite{CAD:hydro} appeared, where the same problem is independently studied for quantum field theories with diagonal scattering and a generalised hydrodynamical description is developed. In this framework, the authors derive an equation analogous to \eqref{Eq:continuity}.
\end{acknowledgments}

\pagebreak 

\onecolumngrid

\begin{center}
{\large{\bf Supplemental Material for\\ ``Transport in Out-of-Equilibrium XXZ Chains: Exact Profiles of Charges and Currents''}}
\end{center}

We collect here the technical tools used in the manuscript. 
\begin{itemize}
\item[-] Section \ref{sec:numerics} presents the numerical method used to simulate the XXZ spin chain after a quantum quench from an inhomogeneous state. 
\item [-] Section \ref{tba} is a concise introduction to the thermodynamic Bethe Ansatz (\textsc{tba}) formalism, which is the one we adopted to work out our results. It does \emph{not} aim at being a complete review of the \textsc{tba} and, for a more systematic description,  we refer the reader to \cite{Takahashibook}. Importantly, subsection \ref{excitations} explains the meaning of quasi-particle excitation on a thermodynamic Bethe state, which constitutes the key element of the kinetic theory introduced in the paper. 
\item[-] Section \ref{sec:indentities} derives the \textsc{tba} identities used in the main text. 
\item[-] Section \ref{add_ex} reports additional examples where the initial state consists of two chains prepared at different temperatures and then joined together.
\item[-] Section \ref{zotos_sec} shows the discrepancy between our result and the conjecture put forward in \cite{Zotos}, which turns out to be a very good approximation despite being inexact.   
\end{itemize}

\section{On the numerical simulations}\label{sec:numerics}%
Static and dynamical properties of the open XXZ spin-${1}/{2}$ 
chain can be efficiently investigated using \textsc{mps}-based algorithms.
The starting point is the matrix-product state representation of the initial state
\be\label{eq:Psi_MPS}
|\Psi\rangle = \sum_{s_{1},\ldots s_{L}}
{\Lambda}_{0}{ \Gamma}^{s_1}_{1}{\Lambda}_{1}
{\Gamma}^{s_2}_{2}{\Lambda}_{2}
\cdots
{\Lambda}_{L-1}{\Gamma}^{s_L}_{L-1}{\Lambda}_{L}
|s_1,s_2,\ldots,s_{L}\rangle.
\ee
Here $L$ is the length of the chain; the physical index $s_{j}$ runs over the local Hilbert space spanned by $\{|\!\!\uparrow_{z}\rangle,|\!\!\downarrow_{z}\rangle\}$; 
${\Gamma}^{s_j}_{j}$ are $\chi_{j-1}\times\chi_{j}$ matrices associated with the $j^{th}$
lattice site; ${\Lambda}_{j}$ are diagonal $\chi_{j}\times\chi_{j}$ matrices with as
entries the singular values corresponding to the bipartition of the system at the $j^{th}$ bond
(namely between site $j$ and site $j+1$).

The representation \eqref{eq:Psi_MPS} holds for \emph{pure} states, but can be straightforwardly generalized 
to a mixed state $\boldsymbol{\rho}$~\cite{MPDO}. It reads as
\be\label{eq:rho_MPDO}
\boldsymbol{\rho} = \sum_{s_{1},s'_{1},\ldots s_{L},s'_{L}}
\tilde{\Lambda}_{0}\tilde{ \Gamma}^{s_1,s'_1}_{1}\tilde{\Lambda}_{1}
\tilde{\Gamma}^{s_2,s'_2}_{2}{\Lambda}_{2}
\cdots
\tilde{\Lambda}_{L-1}\tilde{\Gamma}^{s_L,s'_L}_{L-1}\tilde{\Lambda}_{L}
|s_1,s_2,\ldots,s_{L}\rangle \langle s'_1,s'_2,\ldots,s'_{L}|\,.
\ee
Since the density matrix is positive semidefinite, the matrices in \eqref{eq:rho_MPDO}  
can be decomposed as follows
\be\label{eq:positive_semidef}
\tilde{\Gamma}_{j}^{s_j,s'_j}  =  \sum_{r_j=1}^{K_j}
{\Gamma}_{j}^{s_j,r_j} \otimes ({\Gamma}_{j}^{s'_j,r_j})^{*}, \quad
\tilde{\Lambda}_{j}  =  {\Lambda}_{j} \otimes {\Lambda}_{j},
\ee
where the index $r_{j}$ runs on a local auxiliary space
with dimension $K_{j}$. For a pure state, $K_{j}=1$. 

As a consequence of \eqref{eq:rho_MPDO} and \eqref{eq:positive_semidef}, any operation on mixed states can be reformulated using the standard \textsc{mps} language, 
as an operation on a {\it locally purified state}.
In practice, an auxiliary system ({\it ancilla}) with local Hilbert space of dimension
$K_{j}$ is associated with each local spin, allowing to write 
the locally purified \textsc{mps} state as
\be\label{eq:rho_MPS}
|\Psi_{\boldsymbol{\rho}}\rangle = \sum_{s_{1},r_{1},\ldots s_{L},r_{L}}
{\Lambda}_{0}{ \Gamma}^{s_1,r_1}_{1}{\Lambda}_{1}
{\Gamma}^{s_2,r_2}_{2}{\Lambda}_{2}
\cdots
{\Lambda}_{L-1}{\Gamma}^{s_L,r_L}_{L-1}{\Lambda}_{L}
|s_1,r_1,s_2,r_2,\ldots,s_{L},r_{L}\rangle\,.
\ee
The original mixed state is obtained by tracing out the {\it ancillas},
namely $\boldsymbol{\rho} = {\rm Tr}_{K}|\Psi_{\boldsymbol{\rho}}\rangle \langle\Psi_{\boldsymbol{\rho}}|$. We note that for thermal states, which are the only mixed states we consider here, the local {\it ancilla} can be chosen to be a spin-$1/2$, namely $K_{j} = 2$.

Using the representations \eqref{eq:Psi_MPS} and \eqref{eq:rho_MPS}, we implemented both real and imaginary time-evolution using the
Time Evolving Block Decimation (\textsc{tebd}) procedure \cite{TEBD}.
Specifically, the (imaginary) time is discretized and we use the $2^{nd}$-order Suzuki-Trotter decomposition of the time-step evolution operator,
with real time discretization $dt = 10^{-2}$, and 
imaginary time discretization (inverse-temperature step) $d\beta = 10^{-3}$. 

For initial states that can be represented in terms of pure \textsc{mps} states
(\emph{e.g.} $|\Psi_0\rangle = |\text{N\'eel}\rangle_{L/2}\otimes|\text{Bell}\rangle_{L/2}$),
we only need the real-time evolution operator, which is given by
\be
{\mathcal U}^{\textsc{mps}}_{dt} = \prod_{j\, odd} \exp{(-i \hat h_{\textsc{xxz}} dt/2)} 
\prod_{j, even} \exp{(-i \hat h_{\textsc{xxz}} dt)} 
\prod_{j\, odd} \exp{(-i \hat h_{\textsc{xxz}} dt/2)}.
\ee
Here $\hat h_{\textsc{xxz}} = {\boldsymbol s}^{x}\otimes {\boldsymbol s}^{x}+{\boldsymbol s}^{y} \otimes {\boldsymbol s}^{y} + \Delta\,{\boldsymbol s}^{z} \otimes {\boldsymbol s}^{z}$
is the local XXZ Hamiltonian density.

For thermal states we implemented both real-time and imaginary-time evolution operators
\begin{align}
{\mathcal U}^{\textsc{mpdo}}_{dt} &= \prod_{j\, odd} \exp{(-i \hat h_{r} dt/2)} 
\prod_{j, even} \exp{(-i \hat h_{r} dt)} 
\prod_{j\, odd} \exp{(-i \hat h_{r} dt/2)}\\
{\mathcal R}^{\textsc{mpdo}}_{d\beta} &= \prod_{j\, odd} \exp{(-\hat h_{i} d\beta/2)} 
\prod_{j, even} \exp{(-\hat h_{i} d\beta)} 
\prod_{j\, odd} \exp{(-\hat h_{i} d\beta/2)}.
\end{align}
Here we defined $
\hat h_{r} = \hat h_{\textsc{xxz}} \otimes \hat {\mathbb{I}} - \hat {\mathbb{I}} \otimes\hat h_{\textsc{xxz}}$ and $
\hat h_{i} = \hat h_{\textsc{xxz}} \otimes \hat {\mathbb{I}}$, where
 the operators on the left of the tensor product act on the physical spins and the ones on the right   
on the {\it ancillas}.

The imaginary-time evolution is used to construct thermal states at generic inverse temperature $\beta$ as follows. The infinite temperature state~$| \boldsymbol{\mathbb{I}}\rangle$ admits an exact \textsc{mpdo} representation with auxiliary dimension $\chi_{j}=1$ and $K_j=2$. A \textsc{mpdo} representation of the finite temperature state $|\Psi_{\boldsymbol\rho_\beta}\rangle$ at inverse temperature $\beta$ is then constructed by evolving~$|\boldsymbol{\mathbb{I}}\rangle$ in imaginary time up to inverse temperature $\beta = M d\beta$: 
$
|\Psi_{\boldsymbol\rho_\beta}\rangle =\left({\mathcal R}^{\textsc{mpdo}}_{d\beta}\right)^{M} |\boldsymbol{\mathbb{I}}\rangle.
$
During the imaginary time evolution, $K_j$ remains fixed while the auxiliary dimension $\chi_{j}$ is dynamically
updated by retaining all singular values larger than zero. This can be easily accomplished since 
the temperature introduces a finite correlation length and, in the range of temperatures considered, 
the \textsc{mpdo}
description of the state is free from truncation errors.
The only unavoidable source of error is the Suzuki-Trotter approximation, giving an error of the order $\sim d\beta^{2} = 10^{-6} $.

We followed an analogous procedure also for the real-time evolution, updating the auxiliary dimension dynamically. 
However, we set a maximum auxiliary dimension $\chi_{\textsc{max}}\in[200,400]$ depending on the post-quench value of $\Delta$. With this choice, in the  time window explored ($t\in[0,20]$), the total error remains $\sim 10^{-4}$.

In the numerical implementation, we partially fixed the gauge freedom of the \textsc{mps} by imposing both the left and right canonical form; namely the tensors satisfy the following relations
\be
\sum_{s_{j},r_{j}} (\Gamma_{j}^{s_{j},r_{j}})^{*}\Lambda_{j-1}^{2} \Gamma_{j}^{s_{j},r_{j}} 
= \mathbb{I}_{\chi_{j}},\quad
\sum_{s_{j},r_{j}} \Gamma_{j}^{s_{j},r_{j}}\Lambda_{j}^{2} (\Gamma_{j}^{s_{j},r_{j}})^{*} 
= \mathbb{I}_{\chi_{j-1}},
\ee
with $\mathbb{I}_{\chi}$ the $\chi \times \chi$ identity matrix.
As a consequence, the expectation value of any local quantity $\mathcal{O}_{[m,n]}$
with support in the lattice interval $[m,n]$ (with $n>m$), is given by
\be
\langle \mathcal{O}_{[m,n]} \rangle = 
\langle s_m,\ldots,s_{n}|\mathcal{O}_{[m,n]}|s'_m,\ldots,s'_{n}\rangle
{\rm Tr}\left[
\Lambda_{n}(\Gamma^{s_n,r_n}_{n})^{*}
\cdots
(\Gamma^{s_m,r_m}_{m})^{*}\Lambda_{m-1}
\Lambda_{m-1}\Gamma^{s'_m,r_m}_{m}
\cdots
\Gamma^{s'_n,r_n}_{n}\Lambda_{n}\right],
\ee
where implicit summation over the repeated indices is understood. 
\subsection{Charges and currents}
We implemented the charge densities:
\be
{\boldsymbol s}^{z}_\ell,
\ee
\be
{\mathbf q}^{(1/2)}_{1,\ell} = 
 {\boldsymbol s}^{x}_\ell {\boldsymbol s}^{x}_{\ell+1}+{\boldsymbol s}^{y}_\ell  {\boldsymbol s}^{y}_{\ell+1} + \Delta\,{\boldsymbol s}^{z}_\ell  {\boldsymbol s}^{z}_{\ell+1} - \frac{\Delta}{4},
\ee
\be
{\mathbf q}^{(1/2)}_{2,\ell}  =   {\boldsymbol s}^{x}_{\ell-1}{\boldsymbol s}^{z}_\ell{\boldsymbol s}^{y}_{\ell+1} - {\boldsymbol s}^{y}_{\ell-1}{\boldsymbol s}^{z}_\ell{\boldsymbol s}^{x}_{\ell+1} -   \Delta\, {\boldsymbol s}^{z}_{\ell-1}{\boldsymbol s}^{x}_\ell{\boldsymbol s}^{y}_{\ell+1} + \Delta\, {\boldsymbol s}^{z}_{\ell-1}{\boldsymbol s}^{y}_\ell{\boldsymbol s}^{x}_{\ell+1} -  \Delta\, {\boldsymbol s}^{x}_{\ell-1}{\boldsymbol s}^{y}_\ell{\boldsymbol s}^{z}_{\ell+1} + \Delta\, {\boldsymbol s}^{y}_{\ell-1}{\boldsymbol s}^{x}_\ell{\boldsymbol s}^{z}_{\ell+1}\,,
\ee
\begin{multline}
{\mathbf q}^{(1/2)}_{3,\ell}    = 
-\frac{1+\Delta^{2}}{2} {\boldsymbol s}^{x}_{\ell-2}{\boldsymbol s}^{x}_{\ell-1}
-\frac{1+\Delta^{2}}{2} {\boldsymbol s}^{y}_{\ell-2}{\boldsymbol s}^{y}_{\ell-1}
-\Delta \, {\boldsymbol s}^{z}_{\ell-2}{\boldsymbol s}^{z}_{\ell-1} 
 +  \frac{\Delta}{2} {\boldsymbol s}^{x}_{\ell-2} {\boldsymbol s}^{x}_\ell
+  \frac{\Delta}{2} {\boldsymbol s}^{y}_{\ell-2} {\boldsymbol s}^{y}_\ell
+  \frac{\Delta}{2} {\boldsymbol s}^{z}_{\ell-2}{\boldsymbol s}^{z}_\ell \\
+ 2\Delta \, {\boldsymbol s}^{x}_{\ell-2}{\boldsymbol s}^{y}_{\ell-1}{\boldsymbol s}^{x}_\ell{\boldsymbol s}^{y}_{\ell+1}
+ 2\Delta \, {\boldsymbol s}^{x}_{\ell-2}{\boldsymbol s}^{z}_{\ell-1}{\boldsymbol s}^{x}_\ell{\boldsymbol s}^{z}_{\ell+1}
 +  2\Delta \, {\boldsymbol s}^{y}_{\ell-2}{\boldsymbol s}^{x}_{\ell-1}{\boldsymbol s}^{y}_\ell{\boldsymbol s}^{x}_{\ell+1}
+ 2\Delta \, {\boldsymbol s}^{y}_{\ell-2}{\boldsymbol s}^{z}_{\ell-1}{\boldsymbol s}^{y}_\ell{\boldsymbol s}^{z}_{\ell+1} \\
 +  2\Delta \, {\boldsymbol s}^{z}_{\ell-2}{\boldsymbol s}^{x}_{\ell-1}{\boldsymbol s}^{z}_\ell{\boldsymbol s}^{x}_{\ell+1}
+ 2\Delta \, {\boldsymbol s}^{z}_{\ell-2}{\boldsymbol s}^{y}_{\ell-1}{\boldsymbol s}^{z}_\ell{\boldsymbol s}^{y}_{\ell+1} 
 -  2\Delta \, {\boldsymbol s}^{x}_{\ell-2}{\boldsymbol s}^{y}_{\ell-1}{\boldsymbol s}^{y}_\ell{\boldsymbol s}^{x}_{\ell+1}
- 2\Delta \, {\boldsymbol s}^{y}_{\ell-2}{\boldsymbol s}^{x}_{\ell-1}{\boldsymbol s}^{x}_\ell{\boldsymbol s}^{y}_{\ell+1}\\
 -  2\, {\boldsymbol s}^{x}_{\ell-2}{\boldsymbol s}^{z}_{\ell-1}{\boldsymbol s}^{z}_\ell{\boldsymbol s}^{x}_{\ell+1}
- 2\, {\boldsymbol s}^{y}_{\ell-2}{\boldsymbol s}^{z}_{\ell-1}{\boldsymbol s}^{z}_{\ell}{\boldsymbol s}^{y}_{\ell+1} 
 -  2\Delta^{2} \, {\boldsymbol s}^{z}_{\ell-2}{\boldsymbol s}^{x}_{\ell-1}{\boldsymbol s}^{x}_\ell{\boldsymbol s}^{z}_{\ell+1}
- 2\Delta^{2} \, {\boldsymbol s}^{z}_{\ell-2}{\boldsymbol s}^{y}_{\ell-1}{\boldsymbol s}^{y}_\ell{\boldsymbol s}^{z}_{\ell+1} +  \frac{\Delta}{8}\, ,
\end{multline}
and charge currents:
\be
{\mathbf j}_{\ell}[\boldsymbol{S}^z]   =  {\boldsymbol s}^{x}_{\ell-1}{\boldsymbol s}^{y}_\ell-{\boldsymbol s}^{y}_{\ell-1}{\boldsymbol s}^{x}_\ell  \,,
\ee
\be
{\mathbf j}^{(1/2)}_{1,\ell}  =  {\mathbf q}^{(1/2)}_{2,\ell} ,
\ee
\be
{\mathbf j}^{(1/2)}_{2,\ell}  = \frac{1}{2}  {\mathbf q}^{(1/2)}_{3,\ell} 
+ \frac{1+\Delta^{2}}{2} {\boldsymbol s}^{x}_{\ell-1}{\boldsymbol s}^{x}_{\ell}
+\frac{1+\Delta^{2}}{2} {\boldsymbol s}^{y}_{\ell-1}{\boldsymbol s}^{y}_{\ell}
+\Delta \, {\boldsymbol s}^{z}_{\ell-1}{\boldsymbol s}^{z}_{\ell}-  \frac{\Delta}{4} {\boldsymbol s}^{x}_{\ell-2} {\boldsymbol s}^{x}_{\ell}
-  \frac{\Delta}{4} {\boldsymbol s}^{y}_{\ell-2}{\boldsymbol s}^{y}_{\ell}
-  \frac{\Delta}{4} {\boldsymbol s}^{z}_{\ell-2} {\boldsymbol s}^{z}_{\ell} - \frac{\Delta}{16}\, .
\ee

\section{Thermodynamic Bethe-ansatz for the XXZ spin-$\frac{1}{2}$ chain}\label{tba} %
The XXZ Hamiltonian can be diagonalized by Bethe Ansatz. We take the ferromagnetic state $\left|\uparrow \uparrow \ldots \uparrow \right\rangle =\left|\uparrow\right\rangle^{\otimes L} $ with all spins up as a reference state and build interacting spin waves as excitations over this state. A state with $N$ down spins has total magnetization $\braket{{\boldsymbol S}^z}$ given by $L/2-N$ and is 
characterized by a set of complex rapidities $\{\lambda_j\}_{j=1}^N$. It can be written 
as follows
\begin{subequations}\label{eq:BA_state}
\begin{equation}
	|   \{\lambda_j\}_{j=1}^N \rangle = \sum_{\{ x_j\}} \Psi_{N}\!\left( \{x_j\}_{j=1}^N | \{\lambda_j\}_{j=1}^N  \right)\ \sigma_{x_1}^-\ldots\sigma_{x_N}^-\left|\uparrow\uparrow\ldots\uparrow\right\rangle \epc
\end{equation}
where $x_j$ denotes the position of the $j$-th down spin, and we ordered the sequence so that $x_j<x_k$ for $j<k$. The wave function in coordinate space takes the Bethe Ansatz form
\begin{equation} 
	\Psi_{N}\!\left( \{x_j\}_{j=1}^N | \{\lambda_j\}_{j=1}^N  \right)=\sum_{Q\in\mathcal{S}_N} (-1)^{[Q]} 
 \exp\Bigl\{- i \sum_{j=1}^N x_j\, p(\lambda_{Q_j}) - \frac{i}{2} \sum_{\substack{ j,k=1\\ k>j}}^N \theta(\lambda_{Q_k} - \lambda_{Q_j}) \Bigr\}\epp
\end{equation}
\end{subequations}
The sum runs over the set of all permutations of integers $1, \ldots, N$, denoted by $\mathcal{S}_N$, and $(-1)^{[Q]}$ is the parity of the permutation $Q\in\mathcal{S}_N$. The single particle momentum is given by
\begin{equation}\label{eq:momentum}
 p(\lambda) = - i \ln\left[\frac{\sinh(\lambda + \frac{i \gamma }{2})}{\sinh(\lambda- \frac{i \gamma}{2})}\right]\, .
\end{equation}
The parameter $\gamma $ is such that $\Delta=\cos(\gamma)$. The scattering phase shift $\theta$ reads as
\begin{equation} \label{eq:scateringphaseXXZ}
	\theta(\lambda) =2 \arctan \left( \frac{\tanh(\lambda)}{\tan(\gamma)} \right)\epp
\end{equation}
If the rapidities $\{ \lambda \}_{j=1}^N$ satisfy the Bethe equations,
\begin{equation}\label{eq:BAE_neel}
	\left(\frac{\sinh(\lambda_j+i\gamma/2)}{\sinh(\lambda_j-i\gamma/2)}\right)^L=-\prod_{k=1}^N\frac{\sinh(\lambda_j-\lambda_k+i\gamma)}{\sinh(\lambda_j-\lambda_k-i\gamma)}\:, \qquad j=1,\ldots,N \,,
\end{equation}
the state~\eqref{eq:BA_state} is an eigenstate of the Hamiltonian and is called a ``Bethe state". The solutions to the equations \eqref{eq:BAE_neel} are complex numbers which can be arranged in ``strings'', \emph{i.e.}, they can be parametrized as follows~\cite{Takahashibook}:
\begin{equation}\label{eq:strings}
\lambda_j \to \lambda_{\alpha}^{k} + i \frac{\gamma}{2} (n_k + 1 -2 a)  + i  \frac{\pi (1- \upsilon_k)}{4}  + \delta_{k, \alpha}^a \:, \qquad k=1,\dots, N_s\,,\quad a=1,\ldots,n_k\,,\quad\upsilon_k= \pm 1\, .
\end{equation}
Here $N_s$ denotes the number of the species of excitations, $k$ labels the species, $n_k$ is the length of the corresponding string, $\alpha$ indexes the strings of a given species, and $a$ runs over the rapidities in the same string.  
The index $\upsilon_k$ is called ``parity'' and $-\infty <\lambda_{\alpha}^{k} < + \infty$ are the real \textit{string centers}, which are representative of the real parts of the rapidities in the same string. 
For a given state, in most of the strings the deviations $\delta_{n, \alpha}^a$ vanish exponentially with the system size  $\delta_{n, \alpha}^a \sim e^{- L \alpha_{n, \alpha}^a}$. This is in accordance with the string hypothesis: for large enough $L$,  all the solutions of the Bethe equations are arranged in perfect strings (rapidities disposed as in \eqref{eq:strings} with zero deviations $\delta_{n, \alpha}^a =0$); a state is then solely characterized by the real string centers $\lambda_{\alpha}^k$. The logarithmic form of the Bethe Eqs~\eqref{eq:BAE_neel} can be recast into the Bethe-Gaudin-Takahashi equations for string centers \cite{Takahashibook}
\begin{eqnarray}
L\theta_j(\lambda^j_{\alpha}) - \sum_{k=1}^{N_s} \sum_{\beta=1}^{M_k} \Theta_{jk}(\lambda^j_{\alpha} - \lambda^k_{\beta}) = 2\pi I^j_{\alpha}\, ,
\label{BAE_log_strings}
\end{eqnarray}
where $I^j_{\alpha}$ is integer, if the number of string centers $M_j$ of the $j$-th species is odd, and half-integer, if $M_j$ is even. Note the numbers of string centers are such that 
\begin{equation}
\sum_{k=1}^{N_s} n_k M_k = N \epp
\end{equation}
The dispersion kernels and scattering phases appearing here are 
\begin{align}
\theta_{j} (\lambda) &=2 \upsilon_j ~\mbox{atan} \left[(\tan n_j\gamma/2)^{-\upsilon_j}\tanh \lambda \right] \equiv \theta^{\upsilon_j}_{n_j}  (\lambda), \nonumber \\
\Theta_{jk}(\lambda) &=(1-\delta_{n_j n_k}) \theta^{\upsilon_j \upsilon_k}_{|n_j-n_k|}(\lambda) + 2\theta^{\upsilon_j \upsilon_k}_{|n_j-n_k|+2}(\lambda) + ... + 
2\theta^{\upsilon_j \upsilon_k}_{n_j+n_k-2} (\lambda) + \theta^{\upsilon_j \upsilon_k}_{n_j+n_k} (\lambda)\, .
\end{align}
At the so-called ``roots of unity'' points $\gamma = \frac{\pi}{\nu_1 + \frac{1}{\nu_2 + \frac{1}{\nu_3 + \ldots}}}$ the number of string types is 
\begin{equation}
N_s= \sum_{k} \nu_k \, .
\end{equation}
In the thermodynamic limit, for each string type the rapidities  become dense on the real axis $-\infty< \lambda  < \infty$. Let us introduce the \emph{counting functions} $\{z_j(\lambda)\}_{j=1}^{N_s}$
\be
z_j(\lambda)=\theta_j(\lambda) -\frac{1}{L} \sum_{k=1}^{N_s} \sum_{\beta=1}^{M_k} \Theta_{jk}(\lambda - \lambda^k_{\beta})\,.
\ee
These are monotonic functions such that 
\be
z_j(\lambda_\alpha^j)=\frac{2\pi I_\alpha^j}{L}\,,
\ee
where $\{\lambda_\alpha^j\}$ are the solutions to \eqref{BAE_log_strings}. These relations can be used to establish a one to one correspondence between rapidities and integers
\be
\label{eq:correspondence}
\lambda_{I^j}^j\longmapsto I^j \quad\text{such that}\quad
z_j(\lambda_{I^j}^j)=\frac{2\pi I^j}{L}\,.
\ee
In particular, for a given solution $\{\lambda_\alpha^j\}$ of \eqref{BAE_log_strings}, \eqref{eq:correspondence} produces a corresponding set of integers, which we say to be \emph{occupied}. The other integers $\{\bar I_\alpha^j\}$ corresponding through \eqref{eq:correspondence} to rapidities $\{\bar\lambda_\alpha^j\}$ which do not appear in $\{\lambda_\alpha^j\}$ are said  \emph{empty}. Rapidities corresponding to occupied and empty integers are called \emph{particles} and \emph{holes} respectively. 
Since particles and holes become dense in the thermodynamic limit, one can recast the problem of finding their actual values into the determination of their ``macroscopic'' distributions  $\rho_{j} (\lambda)$ and $\rho_{j}^h (\lambda)$, defined as 
\begin{equation}
\rho_{j} (\lambda^j_\alpha) = \lim_{L \to \infty} \frac{1}{L |\lambda^j_{\alpha + 1} - \lambda^j_\alpha |} \,, \qquad\qquad
\rho^h_{j} (\bar \lambda^j_\alpha) = \lim_{L \to \infty} \frac{1}{L |\bar \lambda^j_{\alpha + 1} - \bar \lambda^j_\alpha |} \,.
\end{equation} 
These functions are called \emph{root densities} and are related to the counting functions as follows
\be
\label{eq:rhotz}
 \rho^{t}_j(\lambda)\equiv\rho_j(\lambda)+\rho_j^h(\lambda)= \frac{\sigma_j }{2 \pi} \frac{{\rm d}}{{\rm d}\lambda}z_j(\lambda)\,,
\ee
where
\begin{equation}
\sigma_j = \text{sign}(q_j)
\label{Eq:signqj}
\end{equation}
and $q_j$ is an auxiliary integer variable which will be defined later for the cases investigated.
 
For later reference, it is useful to introduce the following notation 
\be
\eta_j(\lambda)=\frac{\rho_j^{h}(\lambda)}{\rho_j(\lambda)},\qquad\qquad
\vartheta_j(\lambda)=\frac{\rho_j(\lambda)}{\rho_j^{t}(\lambda)}.
\ee

In the thermodynamic limit, equations \eqref{BAE_log_strings} involve both counting functions and root densities. The dependence on counting functions can be removed by differentiating and making use of \eqref{eq:rhotz}; this leads  to the so called \textsc{tba} equations
\be
a_j(\lambda)=\sigma_j \left[\rho_j(\lambda)+\rho_j^h(\lambda)\right]+\sum_{k}\int_{-\infty}^{\infty}{\rm d}\mu\,T_{jk}(\lambda-\mu)\rho_k(\mu)\,.
\label{eq:tba}
\ee
Here the sum over $k$ runs over all the string types and we introduced 
\begin{align}
a_j  (\lambda) &= \frac{1}{2 \pi} \frac{{\rm d}}{{\rm d}\lambda} \theta_{j} (\lambda)= \frac{\upsilon_j}{\pi} \frac{\sin(\gamma n_j)}{\cosh(2\lambda) - \upsilon_j \cos(\gamma n_j)}\equiv a^{\upsilon_j}_{n_j}  (\lambda) \epc \\
T_{jk}(\lambda)  &=\frac{1}{2 \pi} \frac{{\rm d}}{{\rm d}\lambda} \Theta_{jk}(\lambda) =(1-\delta_{n_j n_k}) a_{|n_j-n_k|}^{\upsilon_j \upsilon_k}(\lambda) + 2a_{|n_j-n_k|+2}^{\upsilon_j \upsilon_k}(\lambda) + ... + 
2a_{n_j+n_k-2}^{\upsilon_j \upsilon_k} (\lambda) + a_{n_j+n_k}^{\upsilon_j \upsilon_k} (\lambda)\,.
\end{align}
The distributions $\rho_j$ are normalized by the total number of particles, which is in a simple relation to the magnetization density
\begin{equation}
\frac{N}{L}=L^{-1} \sum_{k,\beta}   1  \xrightarrow{L\rightarrow\infty}
\sum_{k} \int d\lambda\ \rho_k(\lambda) n_k = \frac{1}{2} - \frac{1}{L}\langle \rho | \boldsymbol{S}^z | \rho \rangle  \epp
\end{equation}
The energy density $E = \frac{1}{L}\langle \rho  | \boldsymbol{H} | \rho \rangle $  is instead given by 
 \begin{equation}
E=  L^{-1} \sum_{j=1}^N e(\lambda_j) + \frac{\Delta}{4} =  L^{-1} \sum_{k,\beta} e_k(\lambda_\beta^k)  + \frac{\Delta}{4}   \xrightarrow{L\rightarrow\infty} \sum_{k} \int d\lambda \ \rho_k(\lambda) e_k  (\lambda) + \frac{\Delta}{4}\,, 
\end{equation}
where $e_k  (\lambda)$ is the energy of a string of type $k$ with rapidity $\lambda$, obtained by summing over all the contributions of the single particles inside the string
\begin{equation}
 e_k  (\lambda) = \sum_{a=1}^{n_k}   e \left(\lambda + i \frac{\gamma}{2} (n_k + 1 -2 a) + i \frac{(1- \upsilon_k) \pi }{4} \right)\,.
\end{equation}
Equivalently, any other charge density of a thermodynamic state can be expressed in terms of the distribution of particles
 \begin{equation}
\frac{1}{L}\langle \rho  | \boldsymbol{Q}^{(s)}_n | \rho \rangle=  L^{-1} \sum_{j=1}^N q^{(s)}_n(\lambda_j) =  L^{-1} \sum_{k,\beta}  q^{(s)}_{n,k}(\lambda_\beta^k)    \xrightarrow{L\rightarrow\infty} \sum_{k} \int d\lambda \ \rho_k(\lambda) q^{(s)}_{n,k}  (\lambda)\,,
\end{equation}
where $q^{(s)}_n(\lambda) $ is the single particle eigenvalue and
\begin{equation}
q^{(s)}_{n,k}  (\lambda) = \sum_{a=1}^{n_k}  q^{(s)}_{n}  \left(\lambda + i \frac{\gamma}{2} (n_k + 1 -2 a) + i \frac{(1- \upsilon_k) \pi }{4} \right)\,. 
\end{equation}

\subsection{Excitations over a generic thermodynamic state}\label{excitations}%

A thermodynamic state $| \rho \rangle$  can be defined as the thermodynamic limit of a single representative state specified by a set of quantum numbers   $ \{ I_\alpha^j\,,\,\,\alpha=1,\ldots, M_j,\,\,j=1,\ldots,N_s \}$ which corresponds to a set of rapidities $ \{ \lambda_\alpha^j\,,\,\,\alpha=1,\ldots, M_j,\,\,j=1,\ldots,N_s \}$. 
At fixed string type $j$, we can define excitations on top of this state, as displacements of some quantum numbers $I^j_\alpha \to I'^j$. We may then split the set of rapidities in ``non-excited ones" $\{ \tilde{\lambda}_\alpha^j\}_{\alpha=1}^{M_j-m_j}$, with $m_j \ll M_j$, a set of particle excitations  $\{ \lambda_{\alpha}^{+\,j} \}$, and a set of hole excitations $\{ \lambda^{-\,j}_{\alpha} \}$. The latter contains fictitious rapidities which represent the empty slots left by the particle excitations. The total number of excitations is assumed to be even. 
The rapidities in the first set $\{ \tilde{\lambda}_\alpha^j\}_{\alpha=1}^{M_j-m_j}$ are related to the ones of the representative state $\{ {\lambda}_\alpha^j\}_{\alpha=1}^{M_j-m_j}$  as follows  
\begin{equation}\label{eq:deviationex}
\tilde{\lambda}_\alpha^j =  {\lambda}_\alpha^j  + \sum_{k} \sum_{\beta=1}^{n_{exc}^{k}} \frac{s_\beta F_{jk}(\lambda_\alpha^j| \lambda^{s_\beta\,k} _{\beta})}{\sigma_j \rho^t_j(\lambda_\alpha^j) L} + \mathcal{O}(L^{-2}) \, .
\end{equation}
where $s_\beta$ is equal to $+$ ($-$) for particle (hole) excitations,
$n_{exc}^{k}$ is the number of excitations for given species $k$, and  
$F_{jm}(\lambda|\mu)$ is the so-called shift function satisfying the following integral equation
\be\label{eq:shift}
F_{jm}(\lambda|\mu)=\frac{1}{2\pi}\Theta_{jm}(\lambda-\mu) - \sum_{k}\int{\rm d}\kappa T_{jk}\left(\lambda - \kappa \right)\vartheta_k(\kappa)\sigma_k F_{km}(\kappa | \mu)\,. 
\ee
Given equation \eqref{eq:deviationex} we can then write down the corrections to the the energy of the state due to the presence of the particle-hole excitations 
\begin{align}
\Delta E = &  \left( \sum_{j=1}^{N_s}\sum_{\alpha=1}^{M_j-m_j} e_j(\tilde{\lambda}^j_\alpha) -  e_j(   {\lambda}^j_\alpha) + \sum_{k,\beta} 
s_\beta e_k(\lambda^{s_\beta\,k} _{\beta})  \right) \\& \xrightarrow{L\rightarrow\infty}   \sum_{k,\beta} s_\beta\left(
 e_k(\lambda^{s_\beta \,k} _{\beta}) + \sum_l \int d\lambda \  e_l'(\lambda) F_{l k }(\lambda | \lambda^{s_\beta\,k} _{\beta}) \vartheta_l(\lambda)  \sigma_l\right)\,.
\end{align}
The same computation can be carried out for the difference of momenta
\begin{align}
\Delta P &  =   \left( \sum_{j=1}^{N_s}\sum_{\alpha=1}^{M_j-m_j} \theta_j(\tilde{\lambda}^j_\alpha) -  \theta_j(   {\lambda}^j_\alpha)+ \sum_{k,\beta}s_\beta 
\theta_k(\lambda^{s_\beta\,k} _{\beta})  \right) \\ &  \xrightarrow{L\rightarrow\infty}  \sum_{k,\beta} s_\beta\left(
\theta_k(\lambda^{s_\beta\, k} _{\beta})  + \sum_l \int d\lambda \  \theta_l'(\lambda) F_{l k }(\lambda | \lambda^{s_\beta\,k} _{\beta} ) \vartheta_l(\lambda) \sigma_l \right)\,.
\end{align}

We can then define the energy and momenta for a single particle excitation with rapidity $\mu$ and string type $k$ as
\begin{equation}
\varepsilon_k(\mu)  =  
e_k(\mu)   + \sum_l \int d\lambda \  e_l'(\lambda) F_{l k }(\lambda | \mu )  \vartheta_l(\lambda)   \sigma_l\,,\label{eq:ed}
\end{equation}
\begin{equation}
p_k(\mu)  =  
\theta_k(\mu)   + \sum_l \int d\lambda \  a_l(\lambda) F_{l k }(\lambda | \mu )  \vartheta_l(\lambda) \sigma_l\,. \label{eq:pd}
\end{equation}
These are the so-called dressed energy and momentum. 

This analysis can be easily extended to all the other conserved charges of the model. It turns out that an excitation with rapidity $\mu$ and string type $k$ ``carries'' a dressed charge $\boldsymbol{Q}_n^{(s)}$ given by
\begin{equation}
q^{d\,(s)}_{n,k}(\mu)  =  
q^{(s)}_{n,k}(\mu) + \sum_l \int d\lambda \  q^{(s)\,\prime}_{n,l}(\lambda) F_{l k }(\lambda | \mu )  \vartheta_l(\lambda) \sigma_l\,. \label{eq:qd}
\end{equation}
Here $q^{(s)}_{n,k}(\lambda)$ is the single particle eigenvalue of $\boldsymbol{Q}_n^{(s)}$.

\subsection{Thermal states} %

A special class of thermodynamic Bethe states is given by the \emph{thermal states}, which are equivalent to the following density matrix
\be
\boldsymbol{\rho}=\frac{e^{-\beta (\boldsymbol{H}-h\boldsymbol{S}^z)}}{Z}\,,\qquad\qquad Z=\tr{}{e^{-\beta (\boldsymbol{H}-h\boldsymbol{S}^z)}}\,.
\ee
Here $\beta$ is the inverse temperature and we introduced an external magnetic field $h$ along the $z$ direction.

The thermal states are defined by the following integral equations for the functions $\eta_j = \rho^h_j/\rho_j$
\begin{equation}
\label{eq:therm}
\ln (\eta_j(\lambda)) = \beta(2 \, n_j h   +  e_j(\lambda) )+ \sum_{k}\sigma_k \, \int_{-\infty}^{\infty}\mathrm d\mu T_{jk}(\lambda-\mu) \ln \left( 1 + \eta_{k}^{-1}(\mu) \right) \epp
\end{equation}
Remarkably, for these states (and these states only) the dressed energy is related to the functions $\eta_j$ as 
\begin{equation}
\varepsilon_j(\lambda) = \frac{1}{\beta } \log( \eta_j(\lambda)) \epp
\end{equation}

\subsubsection{String content at  $\gamma = \frac{\pi}{\ell}$}%

 At the simple roots of unity points $\gamma = \pi/\ell$ there are $\ell$ strings with the following lengths $n_j$ and parities $\upsilon_j$
\begin{align}
n_j &= j \epc \qquad \upsilon_j=1 \epc \qquad j=1,2\ldots,\ell-1 \epc \\
n_\ell & = 1 \epc \qquad \upsilon_\ell=-1 \epp
\end{align}
In this case,  the auxiliary integers  $q_j$ (\emph{cf}.~\eqref{Eq:signqj}) read as
\begin{align}
q_j &= \ell -n_j \qquad j=1,2\ldots,\ell-1 \epc \\
q_\ell &= -1   \epp
\end{align}

\subsubsection{String content at $\gamma = \frac{\pi}{\nu_1 + \frac{1}{\nu_2}}$}  %

The more involved case of $\gamma/\pi=1/(\nu_{1}+1/\nu_{2})$ has $\nu_1 + \nu_2$ string types. Here we have:
\begin{align}
&n_{j}=
\begin{cases}
   j        &  1\leq j \leq \nu_{1} - 1 \\
   1+(j-\nu_{1})\nu_{1} & \nu_{1}\leq j \leq \nu_{1}+\nu_{2}-1 \\
   \nu_{1} & j=\nu_1 + \nu_2\,.
\end{cases}\\
&\upsilon_{j}=
\begin{cases}
   j        &  1\leq j \leq \nu_{1} - 1 \\
   -1     &   j=\nu_{1}  \\
\exp{(i \pi \ \text{floor}[ (n_{j}-1) \frac{\nu_{2} }{1+\nu_{1}\nu_{2}} ])} &   \nu_{1} +1 \leq j \leq  \nu_1+\nu_2\,.
\end{cases}\\
&q_j = \begin{cases}
   \frac{1+\nu_{1}\nu_{2}}{\nu_{2}}-j        &  1\leq j \leq \nu_{1} - 1 \\
   \frac{1}{\nu_{2}}(j-\nu_{1})-1 & \nu_{1}\leq j \leq \nu_{1}+\nu_{2}-1 \\
   \frac{1}{\nu_{2}} & j= \nu_{1}+\nu_{2}\,.
\end{cases}
\end{align}

\section{Proof of some identities}\label{sec:indentities}  %

In this section we prove some useful identities. In doing that, it is convenient to introduce the following compact notations. Any function $w_{j}(\lambda)$ of the rapidities $\lambda$ with a string index $j$ is represented by a vector $\vec w$
\be
\label{eq:vec}
[ \vec w ]_{j}(\lambda)= w_{j}(\lambda)\,.
\ee
Any function $A_{jk}(\lambda,\mu)$ of two rapidities with two string indices is instead represented by an operator $\hat{A}$ acting as follows
\be
[{\hat{A}} \vec w ]_{j}(\lambda)=\sum\nolimits_{k}\int{{\rm d}\mu}\, A_{jk}(\lambda,\mu) w_k(\mu)\,.
\ee
The inverse of  $\hat{A}$ is the operator $\hat A^{-1}$ satisfying
\be
\sum\nolimits_{k}\int{{\rm d}\nu}\, [A^{-1}]_{i k}(\lambda,\nu)A_{kj}(\nu,\mu)=\delta(\lambda-\mu)\delta_{i j}\, .
\ee
Thus, the distribution $\delta(\lambda-\mu)\delta_{i j}$ corresponds to the identity $\hat 1$. It is also convenient to define \emph{diagonal} operators $\hat w$ associated with a function of a single rapidity
\be
[\hat w]_{i j}(\lambda,\mu)=\delta(\lambda-\mu)\delta_{i j}w_i(\lambda)\, .
\ee
Finally, the scalar product is defined as 
\be
\label{eq:prod}
\vec v\cdot \vec w=\sum_k\int\mathrm d\lambda\, v_k(\lambda)w_k(\lambda)\, .
\ee

\subsection{Root density} %

In compact notations,  the \textsc{tba} equations \eqref{eq:tba} read as
\be\label{eq:tbaO}
\vec a=\sg\vec\rho^{\, t}+\T\vec\rho\, .
\ee
By definition, the root density $\vec \rho$ is in a simple relation with $\vec\rho^{\, t}$:
\be
\vec \rho=\hat\vartheta \vec\rho^{\, t}\, .
\ee
We can therefore invert \eqref{eq:tbaO} to get $\vec\rho^{\, t}$ 
\be\label{eq:rhot}
\vec\rho^{\, t}=\hat\vartheta^{-1}(\sg\hat\vartheta^{-1}+\T)^{-1}\vec a
\ee
and, in turn, the root density
\be\label{eq:rho}
\vec\rho=(\sg\hat\vartheta^{-1}+\T)^{-1}\vec a\, .
\ee

\subsubsection{Expectation values of charge densities}  %

Using \eqref{eq:rho},  we can easily express the expectation value of a charge density (Eq.~(3) in the main text) in terms of $\vartheta$
\be
\braket{\rho|{ \boldsymbol q}_{\ell}|\rho}=\vec { q}\cdot \vec\rho=\vec { q}\cdot (\sg \hat\vartheta^{-1}+ \T)^{-1}\vec { a}\,.
\ee

\subsection{Expectation values of charge currents}%

Let us rewrite \eqref{eq:qd} in compact notations
\be
\label{eq:qdrewritten}
\vec q^{\,d\,(s)}_n=\vec q^{\,(s)}_n+\hat F^t \sg\hat\vartheta\vec q^{\,(s)\,\prime}_n\, .
\ee
Here $A^t$ is the transpose of $A$ and $\hat F$ is the shift function (\emph{cf.}~Eq.\eqref{eq:shift}). Inverting \eqref{eq:shift} we find 
\be
\hat F^t = - \frac{\hat\Theta}{2 \pi}\left(\1+\hat\sigma\hat\vartheta \hat T\right)^{-1}\,,
\ee
where $[\hat\Theta ]_{ij}(\lambda,\mu)=\Theta_{ij}(\lambda-\mu)$. Plugging this into \eqref{eq:qdrewritten} gives 
\be
\vec q^{\,d\,(s)}_n=\vec q^{\,(s)}_n - \frac{\hat\Theta}{2 \pi}\left(\hat\sigma\hat\vartheta^{-1}+ \hat T\right)^{-1} \vec q^{\,(s)\,\prime}_n\, .
\ee
Considering the equation in ``components'' $[\,\cdot\,]_j(\lambda)$, differentiating with respect to $\lambda$ and rewriting everything in compact notations, we get  
\be\label{eq:qdp}
\vec q^{\,d\,(s)\,\prime}_n=\hat\sigma\hat\vartheta^{-1}\left(\hat\sigma\hat\vartheta^{-1}+ \hat T\right)^{-1} \vec q^{\,(s)\,\prime}_n\, .
\ee

By definition, the velocity of the elementary excitations is the derivative of the energy with respect to the momentum
\be
\label{eq:velocity}
v_n(\lambda)=\frac{\varepsilon_n^{\prime}(\lambda)}{p_n^{\prime}(\lambda)}\, .
\ee
Using \eqref{eq:qdp} we have
\be
\vec p^{\, \prime}=2\pi \sg \hat\vartheta^{-1}(\sg \hat\vartheta^{-1}+\T)^{-1}\vec a=2\pi \sg\vec\rho^{\, t}\, ;
\ee
in components it reads as
\be
p_n^{\, \prime}(\lambda)=2\pi \mathrm{sgn}(q_n)\rho^{t}_n(\lambda)\, .
\ee
From this relation and \eqref{eq:velocity} it follows
\be
\hat v\vec\rho=\frac{1}{2\pi}\sg\hat\vartheta\vec \varepsilon^{\,\prime}\,.
\ee
Applying \eqref{eq:qdp} to the derivative of the excitation energy gives
\be
\vec \varepsilon^{\, \prime}=\sg \hat\vartheta^{-1}(\sg \hat\vartheta^{-1}+\T)^{-1}\vec e^{\, \prime}
\ee
and hence
\be
\label{eq:vrho}
\hat v\vec \rho=\frac{1}{2\pi}(\sg \hat\vartheta^{-1}+\T)^{-1}\vec e^{\, \prime}\, .
\ee

Up to a constant, the current  written in Eq.~(14) of the main text is therefore given by
\be
\label{eq:Jj}
\braket{\rho|\boldsymbol{j}_{\ell}[\boldsymbol{Q}]|\rho} =\vec q\cdot \hat v\vec \rho=\frac{1}{2\pi}\vec q\cdot(\sg \hat\vartheta^{-1}+\T)^{-1}\vec e^{\, \prime}\, .
\ee
In particular, using $\vec q_{1}^{\,(1/2)}=\vec e = -\pi\sin\gamma \vec a$ and $\vec q_{2}^{\,(1/2)}= -(\sin\gamma/2)\vec q_{1}^{\,(1/2)\,\prime}$ we have  
\be
\label{eq:check1}
\braket{\rho|\boldsymbol{j}_{\ell}[\boldsymbol{Q}_1^{(1/2)}]|\rho} =\vec a \cdot(\sg \hat\vartheta^{-1}+\T)^{-1}\vec q_{2}^{\,(1/2)}=\braket{\rho|\boldsymbol{q}_{2,\ell}^{(1/2)}|\rho}\, ,
\ee
where in the last step we used the symmetry of $\hat T$. Eq.~\eqref{eq:check1} proves $ \braket{\rho|{\boldsymbol j}_\ell[\boldsymbol{Q}_1^{(1/2)}]|\rho}\sim \braket{\rho|{\boldsymbol q}_{2,\ell}^{(1/2)}|\rho}$, as claimed in  the main text. 

\subsubsection{Comparison with Ref.~\cite{CAD:hydro}}%
As mentioned in the main text, in Ref.~\cite{CAD:hydro} an exact expression for the expectation values of current densities was independently obtained for integrable quantum field theories with diagonal scattering. In our notations it reads as
\be\label{eq:CAD}
\braket{ {\boldsymbol j}_{\textsc{iqft}}[{\boldsymbol q}_{\textsc{iqft}}]}=\frac{1}{2\pi }\vec e_{\textsc{iqft}}^{\, \prime}\cdot(\hat\vartheta_{\textsc{iqft}}^{-1}+\hat {T}_{\textsc{iqft}})^{-1}\vec {q}_{\textsc{iqft}}\, ,
\ee
where the subscript $\textsc{iqft}$ stands for integrable quantum field theory and $\vec e_{\textsc{iqft}}$, $\hat\vartheta_{\textsc{iqft}}$, $\hat T_{\textsc{iqft}}$, and $\vec q_{\textsc{iqft}}$ have meanings analogous to the corresponding quantities in XXZ. This expression is the integrable quantum field theory equivalent of \eqref{eq:Jj} and, in fact, can be recast as in Eq.~(14) of the main text.

\subsection{Continuity equation}%

We are now in a position to prove Eq.~(15) of the main text. Since $\xt=x/t$, Eq.~(12) can be rewritten as
\be
\label{eq:contxt}
\xt \partial_\xt \vec \rho_\xt-\partial_\xt (\hat v_{\xt}\vec \rho_\xt)=0\, .
\ee
Using the explicit expressions \eqref{eq:rho} and \eqref{eq:vrho} for $\vec \rho$ and $\hat v\vec \rho$ we find
\be
\xt \partial_\xt (\sg\hat\vartheta_{\xt}^{-1}+\T)^{-1}\vec a-\partial_\xt (\sg \hat\vartheta_{\xt}^{-1}+\T)^{-1}\frac{\vec e^{\, \prime}}{2\pi}=0\,.
\ee
Since only  $\hat \vartheta_{\xt}$ depends on $\xt$, the derivative of $(\sg\hat\vartheta_{\xt}^{-1}+\T)^{-1}$ with respect to $\xt$ is readily obtained
\be
\partial_\xt (\sg\hat\vartheta_{\xt}^{-1}+\T)^{-1}=-(\sg\hat\vartheta_{\xt}^{-1}+\T)^{-1}\sg\partial_\xt (\hat\vartheta_{\xt}^{-1})(\sg\hat\vartheta_{\xt}^{-1}+\T)^{-1}\, .
\ee
Plugging this into \eqref{eq:contxt} gives 
\be
-\xt (\sg\hat\vartheta_{\xt}^{-1}+\T)^{-1}\sg\partial_\xt (\hat\vartheta_{\xt}^{-1})(\sg\hat\vartheta_{\xt}^{-1}+\T)^{-1}\vec a+(\sg\hat\vartheta_{\xt}^{-1}+\T)^{-1}\sg\partial_\xt (\hat\vartheta_{\xt}^{-1})(\sg\hat\vartheta_{\xt}^{-1}+\T)^{-1}\frac{\vec e^{\, \prime}}{2\pi}=0\, .
\ee
This can be rewritten as
\be
(\sg\hat\vartheta_{\xt}^{-1}+\T)^{-1} \sg\hat\vartheta_{\xt}^{-1} (\xt \1-\hat v_{\xt})\partial_\xt (\hat\vartheta_{\xt})\vec\rho_{\xt}^{\, t}=0\, .
\ee
Since $(\sg\hat\vartheta_{\xt}^{-1}+\T)^{-1} \sg\hat\vartheta_{\xt}^{-1}$ is invertible, this equation is equivalent to 
\be\label{eq:contThetaO}
(\xt\1 -\hat v_{\xt})\partial_\xt (\hat\vartheta_{\xt})\vec\rho_{\xt}^{\, t}=0\, ,
\ee
which is exactly Eq.~(15) of the main text. 

An interesting corollary of \eqref{eq:contThetaO} is that the continuity equation can be alternatively expressed in terms of the density of holes $\rho_{\xt,k}^h(\lambda)=\rho_{\xt,k}^t(\lambda)-\rho_{\xt,k}(\lambda)$ as follows
\be
\partial_t\vec \rho_{\xt}^{\, h}+\partial_x\hat v_{\xt}\vec\rho_{\xt}^{\, h}=0\, .
\ee 
This is because the transformation mapping particles into holes maps $\vartheta_{\xt,j}(\lambda)$ into $1-\vartheta_{\xt,j}(\lambda)$ and the velocity in itself; the statement then  follows from the invariance of \eqref{eq:contThetaO} under such transformation.

\section{Two temperatures quench: additional example}\label{add_ex} %

\begin{figure}[h!]
\includegraphics[width=0.35\textwidth]{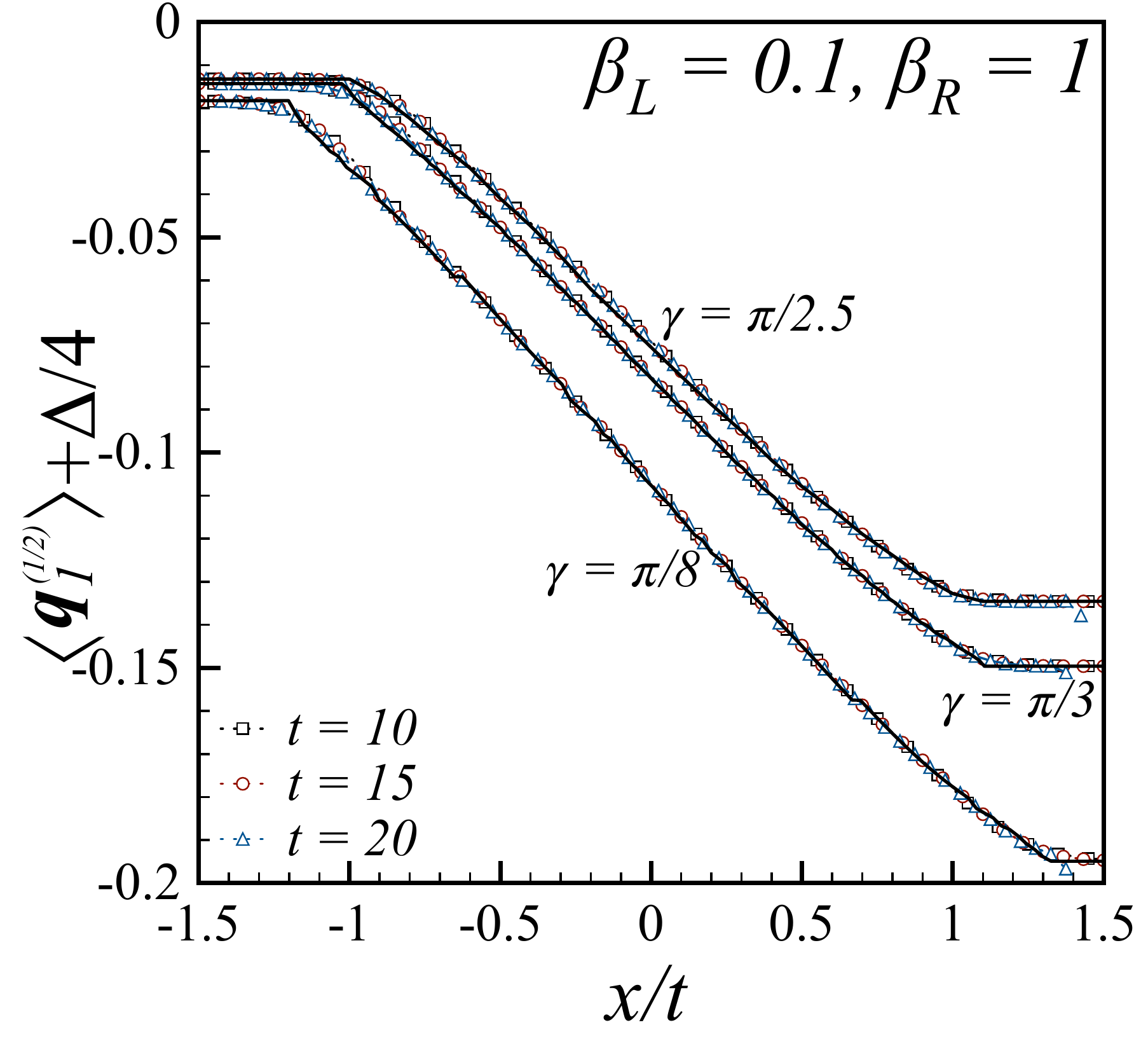}\includegraphics[width=0.35\textwidth]{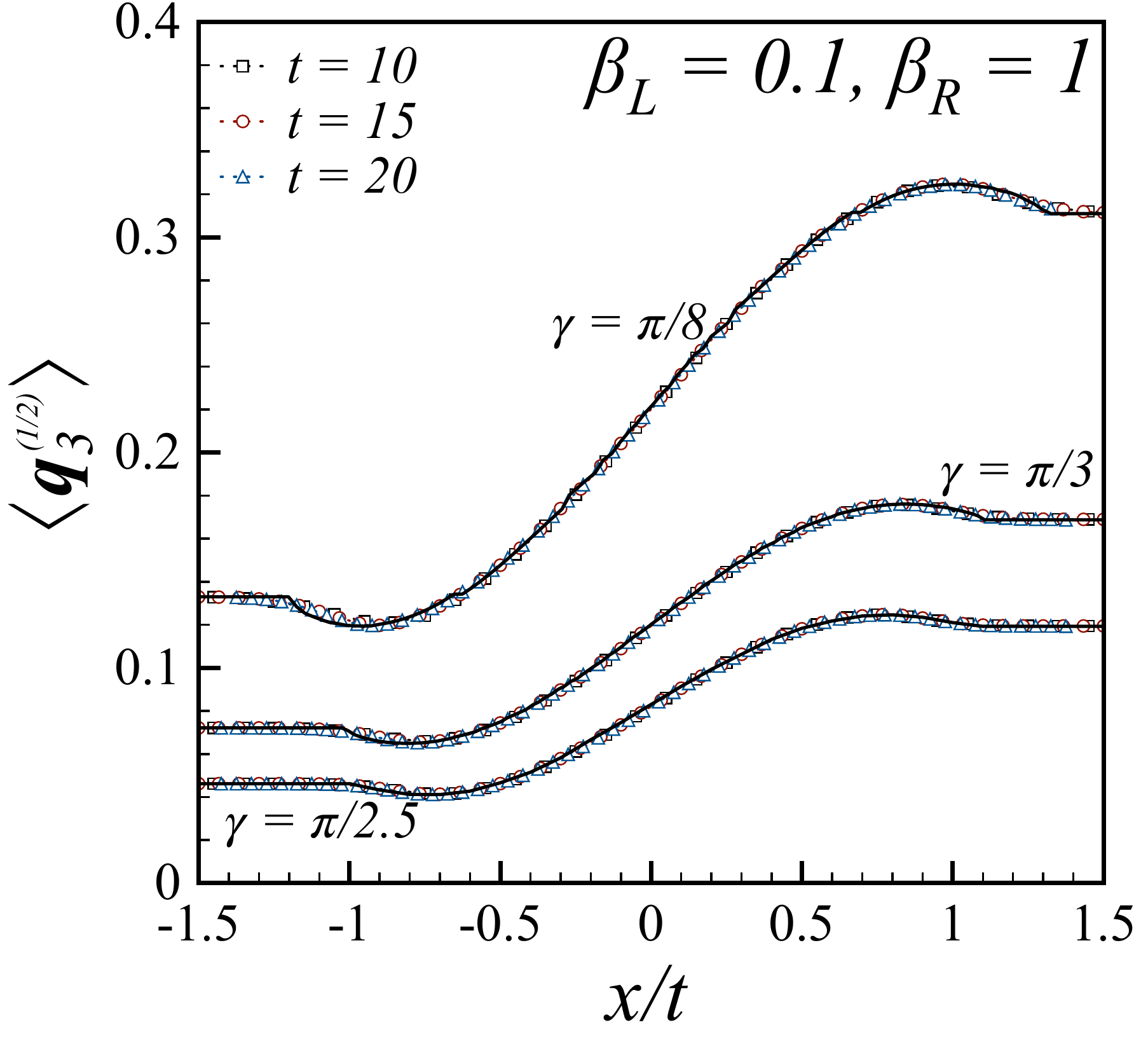} \\
\includegraphics[width=0.35\textwidth]{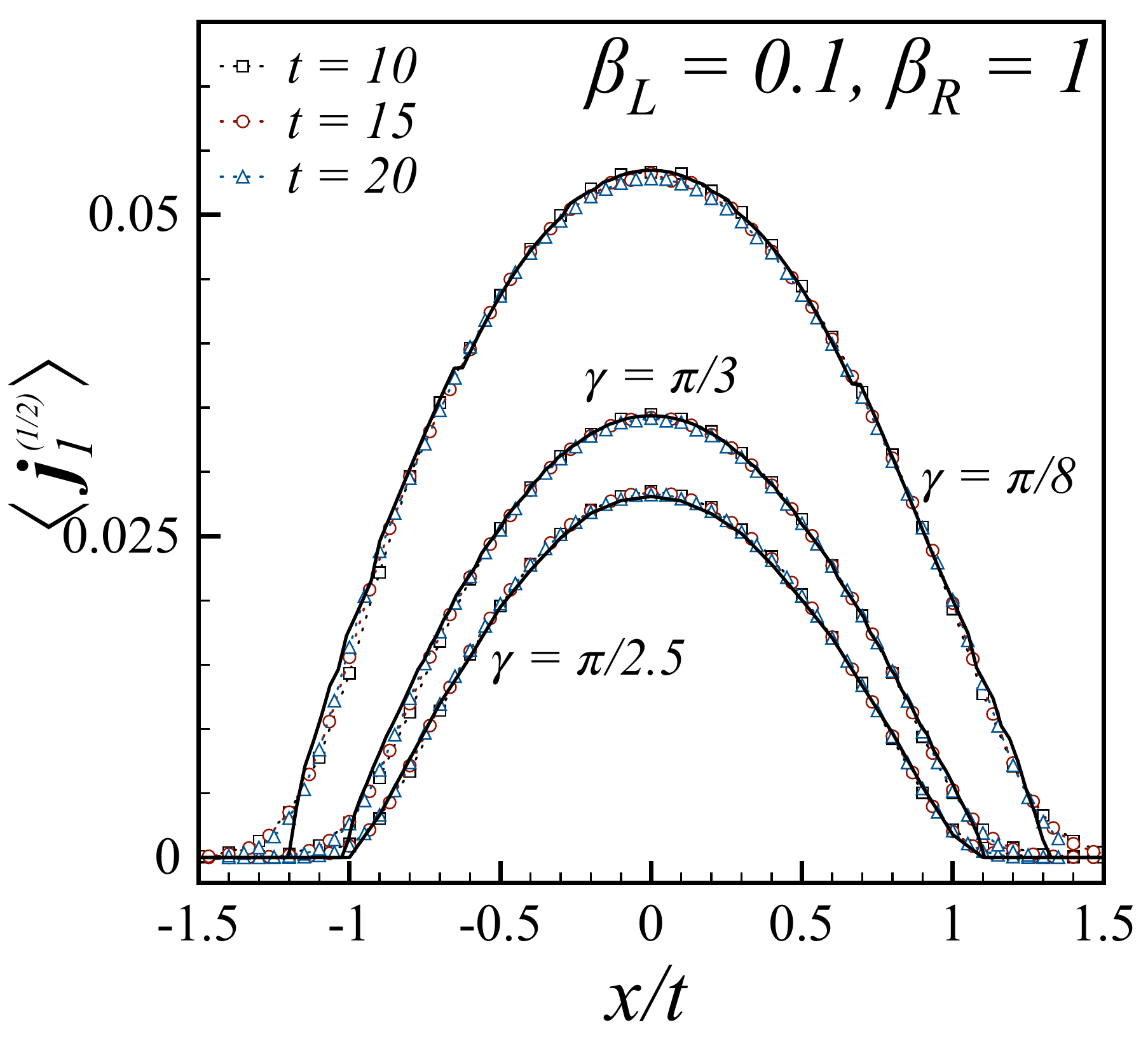}\includegraphics[width=0.35\textwidth]{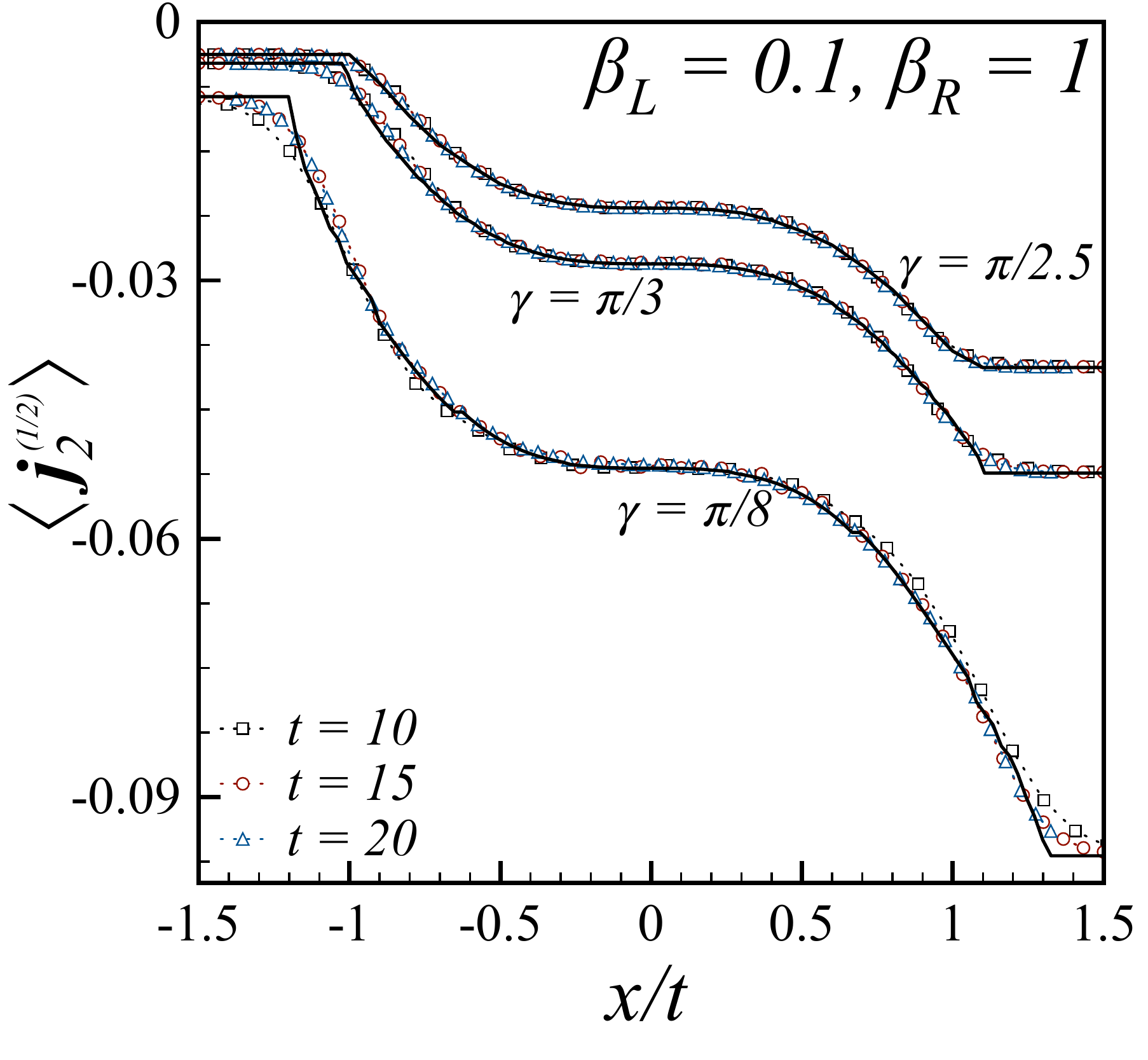}
\caption{Profiles of charge and current densities for three different values of $\Delta = \cos(\gamma)$.  A 60-sites chain has been initially prepared in two halves at inverse temperatures 
$\beta_{L}=0.1$ and $\beta_{R}=1$. Symbols denote 
data obtained via an \textsc{mpdo}-based algorithm; full black lines are the predictions based on (16). 
The tiny ripples in the predictions are numerical artifacts.}.
\label{fig:betaL01_betaR1}
\end{figure}

Here we consider an other example of time evolution after joining two chains at different temperature. In Fig.~\ref{fig:betaL01_betaR1} we report the rescaled profiles of a number of charges and currents ($\boldsymbol{j}^{(1/2)}_{1,\ell}$ is the energy current)
for different times $t=10,\,15,\,20$ and interactions $\Delta$ in the case of $\beta_L=0.1$ and $\beta_R=1$. 

As shown in the figure, the rescaled numerical data are in excellent agreement with the analytical predictions. 
We note that the maximal velocities are smaller than in the case $(\beta_L=1,\beta_R=2)$ reported in Fig.~2 of the main text. 

\section{Comparison with Ref.~\cite{Zotos}}\label{zotos_sec} %

\begin{figure}[h!]
\centering
\includegraphics[width=0.45\textwidth]{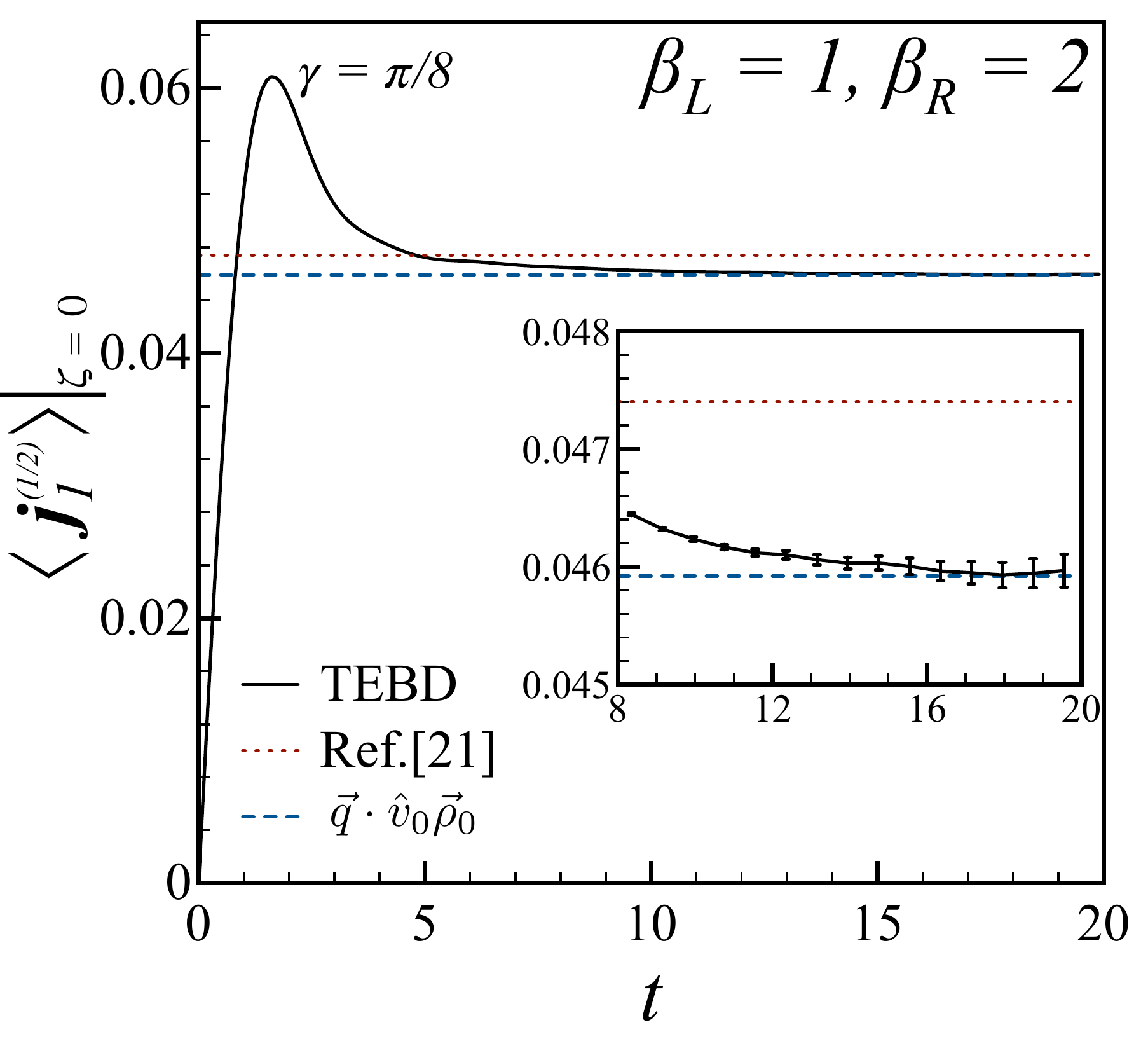}
\hspace{1cm}
\includegraphics[width=0.45\textwidth]{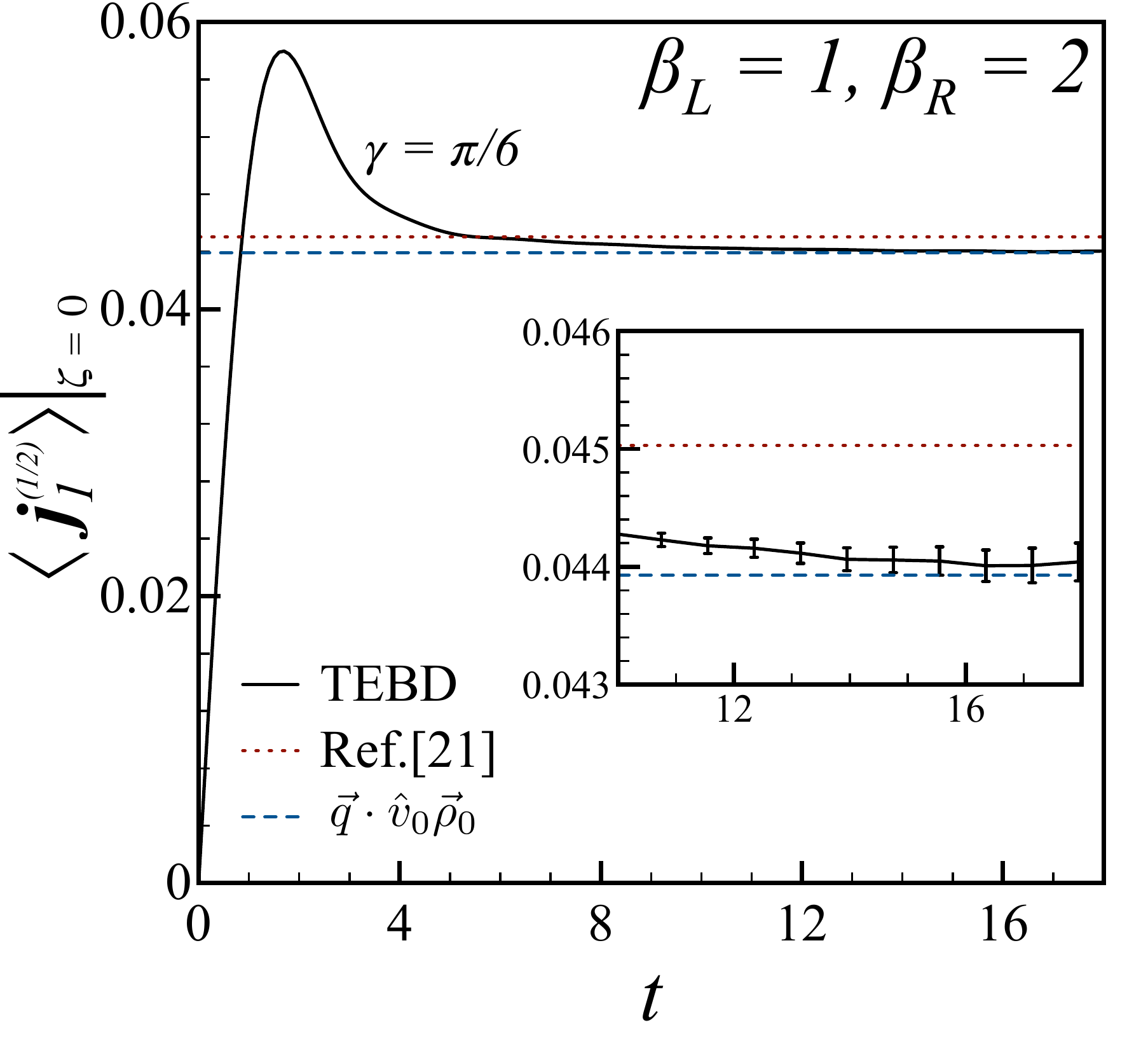}
\caption{Comparison between \textsc{tebd} data, 
the conjecture of Ref.~\cite{Zotos} (left, $0.0474$, and, right, $0.0450$) and our prediction $\vec q\cdot \hat v_0\vec\rho_0$ (left, $0.0459$, and, right, $0.0439$)
for the energy current at $\xt=0$ for $\Delta=\cos(\pi/8)$ (left) and $\Delta=\cos(\pi/6)$ (right). The \textsc{tebd} numerical curve has been obtained by averaging 
the operator ${\mathbf j}^{(1/2)}_{1,\ell}$ over the lattice 
sites $\ell \in \{-1,0,1\}$, in order to smooth-out the lattice effects.
The error bars represent the accumulated truncation error
which remains smaller than $1.5\times 10^{-4}$ for all the explored times.  
The numerical data are consistent with our prediction.\label{fig:comparison}}
\end{figure}

In this section we compare our results with the conjecture of Ref.~\cite{Zotos} for the energy current at $\xt=0$. In compact notations (\emph{cf.}~\eqref{eq:vec}--\eqref{eq:prod}), Ref.~\cite{Zotos} proposed
\be
\braket{\boldsymbol j_{1,\ell}^{(1/2)}}\big|_{\xt=0}=\vec a\cdot \Bigl[\hat \theta_{\textsc{h}}(\sg (\hat\vartheta^{\textsc l})^{-1} +\hat T)^{-1} \hat \theta_{\textsc{h}}+(\hat 1-\hat \theta_{\textsc{h}})(\sg (\hat\vartheta^{\textsc r})^{-1} +\hat T)^{-1} (\hat 1-\hat \theta_{\textsc{h}})\Bigr] \vec q_2^{\, (1/2)}\, .
\ee
Here $\theta_{\textsc{h}}(x)$ is the step function and $\vartheta_j^{\textsc{l(r)}}(\lambda)=\rho^{\textsc{l(r)}}_j(\lambda)/\rho^{t\,\textsc{l(r)}}_j(\lambda)$ where $\rho^{\textsc{l(r)}}_j(\lambda)$ are thermal root densities (\emph{cf.}~\eqref{eq:therm}) at inverse temperature $\beta_{\textsc{l}}$ and $\beta_{\textsc{r}}$ respectively. In Table \ref{t:1} we report a comparison between this conjecture and our result for the same observable, namely $\braket{\boldsymbol j_{1,\ell}^{(1/2)}}|_{\xt=0}=\vec q\cdot \hat v_{0}\vec \rho_0$. We consider $\Delta={1}/{2}$, the value of the anisotropy for which Ref.~\cite{Zotos} showed the best agreement with the numerical simulations.
The differences found are appreciable but smaller than the accuracy of our \textsc{tebd} simulations, see Tab.~\ref{t:1}. In order to distinguish between our prediction and the conjecture of Ref.~\cite{Zotos}, we considered larger values of $\Delta$. The discrepancy between the two results increases with $\Delta$ and the numerical simulations corroborate the exactness of our prediction; see Fig.~\ref{fig:comparison} for two representative examples.

\begin{table}[h!]
\begin{tabular}{c|c|c}
$\beta_R=2\beta_L$&Ref.~\cite{Zotos}&$\vec q\cdot \hat v_{0}\vec \rho_0$\\
\hline
1&0.0191&0.0193\\
2&0.0318&0.0319\\
3&0.0335&0.0333\\
4&0.0294&0.0290\\
5&0.0241&0.0235\\
6&0.0192&0.0186\\
7&0.0153&0.0147\\
8&0.0123&0.0118
\end{tabular}
\vspace{1cm}
\captionof{table}{Comparison between the energy current as conjectured in Ref.~\cite{Zotos} and our result (14) for $\xt=0$ and $\Delta=1/2$.}\label{t:1}
\end{table}

\begin{thebibliography}{99}
\bibitem{f:dr} The reader familiar with  the {\textsc{tba}} could wonder why this expression is not written in terms of dressed quantities. Essentially, this is because a macroscopic number of excitations is moving.
\bibitem{note1} Note that  $u_{\lambda,n}(\xt)\equiv v_{\xt, n}(\lambda)-\xt$ has always an odd number of zeros, indeed the velocity is bounded and hence $\lim_{\xt\rightarrow\mp \infty}u_{\lambda,n}(\xt)= \pm\infty$.
\end{thebibliography}
\end{document}